\documentclass[longauth]{aa}       

\usepackage{savesym}
\usepackage{amsmath}
\savesymbol{iint}
\usepackage[varg]{txfonts}                  
\restoresymbol{TXF}{iint}

\usepackage{graphicx}

\usepackage{url}                 
\usepackage{multirow}            
\usepackage{float}
\usepackage[round]{natbib}       
\usepackage{color}
\usepackage{mathabx}
\newcommand{\acena}{$\alpha \, {\rm Cen\,A}$}          
\newcommand{\acenb}{$\alpha \, {\rm Cen\,B}$}                          %
\newcommand{\acen}{$\alpha \, {\rm Cen}$}                              %

\newcommand{\cmtwo}{cm$^{-2}$}  
\newcommand{\cmthree}{cm$^{-3}$}

\newcommand{\kms}{km\,s$^{-1}$}       

\newcommand{\vlsr}{$\upsilon_{\rm LSR}$}        

\newcommand{\tmb}{$T_{\rm mb}$}

\newcommand{\tsys}{$T_{\rm sys}$}

\newcommand{\ecs}{erg cm$^{-2}$ s$^{-1}$}

\newcommand{\um}{$\mu$m}                                 
\newcommand{\lsun}{$L_{\odot}$}                          
\newcommand{\msun}{$M_{\odot}$}
\newcommand{\rsun}{$R_{\odot}$}

\newcommand{\msunyr}{$M_{\odot} \, {\rm yr}^{-1}$}
\newcommand{\mearth}{$M_{\oplus}$}

\newcommand{\mjup}{$M_{\rm Jupiter}$}
\newcommand{\gapprox}{$\stackrel {>}{_{\sim}}$}   
\newcommand{\lapprox}{$\stackrel {<}{_{\sim}}$}
\newcommand{\about}{$\sim$}                       
\newcommand{\powten}[1]{10$^{#1}$}
\newcommand{\amin}{$^{\prime}$}                   
\newcommand{\asec}{$^{\prime \prime}$}
\newcommand{\adeg}{$^{\circ}$}

\newcommand{\radot}[4]{\mbox{#1$^{\rm h}$#2$^{\rm m}$#3$\stackrel {\rm s}{_{\bf\cdot}}$#4}}  

\newcommand{\decdot}[4]{\mbox{#1$^{\circ}$ #2$^{\prime}$ #3$\stackrel {\prime \prime}{_{\bf \cdot}}$#4}}
\newcommand{\adegdot}[2]{\mbox{#1$\stackrel {\circ}{_{\bf \cdot}}$#2}}
\newcommand{\amindot}[2]{\mbox{#1$\stackrel {\prime}{_{\bf \cdot}}$#2}}
\newcommand{\asecdot}[2]{\mbox{#1$\stackrel {\prime \prime}{_{\bf \cdot}}$#2}}
\newcommand{\ltwo}{$\ell^{\small \rm II}$}
\newcommand{\btwo}{$b^{\small \rm II}$}

\begin{document}
\title{How dusty is \acen tauri?
\thanks{Based on observations with {\it Herschel} which is an ESA space observatory with science instruments provided by European-led Principal Investigator consortia and with important participation from NASA.}
\thanks{and also based on observations with APEX, which is a 12\,m diameter submillimetre telescope at 5100 m altitude on Llano Chajnantor in Chile. The telescope is operated by Onsala Space Observatory, Max-Planck-Institut f\"ur Radioastronomie (MPIfR), and European Southern Observatory (ESO).}}

\subtitle{Excess or non-excess over the infrared photospheres of main-sequence stars}

\author{
                J. Wiegert \inst{1} 
        \and 
                R. Liseau \inst{1}                                              
        \and
                P. Th\'ebault\inst{2}
        \and
                G. Olofsson\inst{3}
        \and
                A. Mora\inst{4}
        \and
                G. Bryden\inst{5}
        \and
                J. P. Marshall\inst{6}
        \and
                C. Eiroa\inst{6}
        \and                                            
                B. Montesinos\inst{7}
        \and            
                D. Ardila\inst{8,\,9}
        \and
                J. C. Augereau\inst{10}
        \and   
                A. Bayo Aran\inst{11,\,12}
        \and
                W. C. Danchi\inst{13}
        \and    
                C. del Burgo\inst{14}
        \and
                S. Ertel\inst{10}
        \and
                M. C. W. Fridlund\inst{15,\,16}
        \and
                M. Hajigholi\inst{1}
        \and
                A. V. Krivov\inst{17}
        \and
                G. L. Pilbratt\inst{18}
        \and
                A. Roberge\inst{19}
        \and
                G. J. White\inst{20,\,21}
        \and
                S. Wolf\inst{22}
}

  \institute{Department of Earth and Space Sciences, Chalmers University of Technology, Onsala Space Observatory, SE-439 92 Onsala, Sweden, 
              \email{wiegert@chalmers.se}  
    \and 
        Observatoire de Paris, Section de Meudon 5, place Jules Janssen, 92195 MEUDON Cedex, Laboratoire d'\'etudes spatiales et d'instrumentation en astrophysique, France
    \and 
        Department of Astronomy, Stockholm University, SE-106 91 Stockholm, Sweden
    \and 
        ESA - ESAC Gaia SOC. P.O. Box 78 E-28691 Villanueva de la Ca{\~n}ada, Madrid, Spain
    \and 
        Jet Propulsion Laboratory, M/S 169-506, 4800 Oak Grove Drive, Pasadena, CA 91109, USA                 
    \and 
        Departamento de F\'{i}sica Te\'{o}rica, C-XI, Facultad de Ciencias, Universidad Aut\'{o}noma de Madrid, Cantoblanco, 28049 Madrid, Spain
    \and 
        Departamento de Astrof\'{\i}sica, Centro de Astrobiolog\'{\i}a (CAB, CSIC-INTA), Apartado 78, 28691 Villanueva de la Ca\~nada, Madrid, Spain
    \and 
        NASA Herschel Science Center, Infrared Processing and Analysis Center, MS 100-22, California Institute of Technology, Pasadena, CA 91125, USA
    \and 
        Herschel Science Center - C11, European Space Agency (ESA), European Space Astronomy Centre (ESAC), P.O. Box 78, Villanueva de la Ca\~nada, 28691 Madrid, Spain
    \and 
        UJF-Grenoble 1 / CNRS-INSU, Institut de Plan\'etologie et d'Astrophysique de Grenoble (IPAG) UMR 5274, Grenoble, F-38041, France 
    \and 
        European Southern Observatory, Casilla 1900, Santiago 19, Chile 
    \and 
        Max Planck Institut f\"ur Astronomie, K\"onigstuhl 17, 69117 Heidelberg, Germany
    \and 
        Astrophysics Science Division, NASA Goddard Space Flight Center, Greenbelt, MD 20771, USA
    \and 
        Instituto Nacional de Astrof{\'i}sica, {\'O}ptica y Electr{\'o}nica, Luis Enrique Erro 1, Sta. Ma. Tonantzintla, Puebla, M{\'e}xico
    \and 
        Institute of planetary Research, German Aerospace Center, Rutherfordstrasse 2, 124 89 Berlin Germany
    \and 
        Leiden Observatory, University of Leiden, P.O. Box 9513, NL-2300 RA Leiden, The Netherlands
    \and 
        Astrophysikalisches Institut und Universit\"atssternwarte, Friedrich-Schiller-Universit\"at Jena, Schillerg\"a\ss chen 2-3, 07745 Jena, Germany 
    \and 
        Astrophysics Mission Division, Research and Scientific Support Department ESA, ESTEC, SRE-SA P.O. Box 299, Keplerlaan 1 NL-2200AG, Noordwijk, The Netherlands 
    \and 
        NASA Goddard Space Flight Center, Exoplanets and Stellar Astrophysics Laboratory, Code 667, Greenbelt, MD  20771, USA
    \and 
        Dept. of Physics \& Astronomy, The Open University, Walton Hall, Milton Keynes MK7 6AA, UK
    \and 
        Space Science \& Technology Department, CCLRC Rutherford Appleton Laboratory, Chilton, Didcot, Oxfordshire OX11 0QX, UK
    \and 
        Institute for Theoretical Physics and Astrophysics, University of Kiel, Leibnizstra{\ss}e 15, D-24098 Kiel, Germany
}

\date{Recieved ... / Accepted ...}


\abstract
{Debris discs around main-sequence stars indicate the presence of larger rocky bodies. The components of the nearby, solar-type binary \acen tauri have higher than solar metallicities, which is thought to promote giant planet formation.}
{We aim to determine the level of emission from debris around the stars in the \acen\ system. This requires knowledge of their photospheres. Having already detected the temperature minimum, $T_{\rm min}$, of \acena\ at far-infrared wavelengths, we here attempt to do so also for the more active companion \acenb.  Using the \acen\ stars as templates, we study possible effects $T_{\rm min}$ may have on the detectability of unresolved dust discs around other stars.}
{We use {\it Herschel-}PACS, {\it Herschel-}SPIRE, and APEX-LABOCA photometry to determine the stellar spectral energy distributions in the far infrared and submillimetre. In addition, we use APEX-SHeFI observations for spectral line mapping to study the complex background around \acen\ seen in the photometric images. Models of stellar atmospheres and of particulate discs, based on particle simulations and in conjunction with radiative transfer calculations, are used to estimate the amount of debris around these stars.}
{For solar-type stars more distant than \acen, a fractional dust luminosity $f_{\rm d} \equiv L_{\rm dust}/L_{\rm star} \sim 2\times 10^{-7}$ could account for SEDs that do not exhibit the $T_{\rm min}$-effect. This is comparable to estimates of $f_{\rm d}$ for the Edgeworth-Kuiper belt of the solar system. In contrast to the far infrared, slight excesses at the $2.5\,\sigma$ level are observed at 24\,\um\ for both \acena\ and B, which, if interpreted to be due to zodiacal-type dust emission, would correspond to $f_{\rm d} \sim (1-3) \times 10^{-5}$, i.e. some $10^2$ times that of the local zodiacal cloud. Assuming simple power law size distributions of the dust grains, dynamical disc modelling leads to rough mass estimates of the putative Zodi belts around the \acen\ stars, viz. \lapprox\,$4 \times 10^{-6}\,M_{\leftmoon}$ of 4 to 1000\,\um\ size grains, distributed according to $n(a)\,\propto\,a^{\,-3.5}$. Similarly, for filled-in $T_{\rm min}$ emission, corresponding Edgeworth-Kuiper belts could account for $\sim 10^{-3}\,M_{\leftmoon}$ of dust.}
{Our far-infrared observations lead to estimates of upper limits to the amount of circumstellar dust around the stars \acena\ and B. Light scattered and/or thermally emitted by exo-Zodi discs will have profound implications for future spectroscopic missions designed to search for biomarkers in the atmospheres of Earth-like planets. The far-infrared spectral energy distribution of \acenb\ is marginally consistent with the presence of a minimum temperature region in the upper atmosphere of the star. We also show that an \acena -like temperature minimum may result in an erroneous apprehension about the presence of dust around other, more distant stars.}

\maketitle

\section{Introduction}

The \acen tauri system lies at a distance of only 1.3\,pc \citep[$\pi=747.1\pm1.2$\,mas,][]{soderhjelm1999}, with the G2\,V star \acena\ (HIP\,71683, HD\,128620) often considered a solar twin. Together with the K\,1 star \acenb\ (HIP\,71681, HD\,128621) these stars are gravitationally bound in a binary system, with  an orbital period of close to 80 years and a semi-major axis (24\,AU) which is intermediate to those of the planets Uranus and Neptune in the Solar system. A third star, Proxima Centauri, about 2\adeg\ southwest of the binary, shares a similar proper motion with them and seems currently to be bound to \acena B, although the M\,6 star \acen\,C (HIP\,70890) is separated by about 15\,000\,AU.

The question as to whether there are also planets around our solar-like neighbours has intrigued laymen and scientists alike. The observed higher metallicities in the atmospheres of \acena\ and B could argue in favour of the existence of planets around these stars \citep[][and references therein]{maldonado2012}. The proximity of \acen\ should allow for highly sensitive observations at high angular resolution with a variety of techniques.

We know today that binarity is not an intrinsic obstacle to planet formation, as more than 12\% of all known exoplanets are seen to be associated with multiple systems \citep{roell2012}. Even if most of these systems have very wide separations ($>100$\,AU) for which binarity might have a limited effect in the vicinity of each star, a handful of planets have been detected in tight binaries of separation $\sim 20$\,AU (e.g., $\gamma$ Cep, HD196885), comparable to that of \acen\ \citep{desidera2007, roell2012}. The presence of these planets poses a great challenge to the classical core-accretion scenario, which encounters great difficulties in such highly perturbed environments \citep[see review in][]{thebault2011}.

For the specific case of \acen tauri, radial velocity (RV) observations indicate that no planets of mass $>2.5$\,\mjup\ exist inside 4\,AU of each star \citep{endl2001}. For their part, theoretical models seem to indicate that in situ planet formation is indeed difficult in vast regions around each star; the outer limit for planet accretion around either star being $\sim 0.5-0.75$\,AU in the most pessimistic studies \citep{thebault2009} and $\sim1-1.5$\, AU in the most optimistic ones \citep[e.g.,][]{xie2010, paardekooper2010}. However, these estimates open the possibility that planet formation should be possible in the habitable zone (HZ) of \acenb, which extends between 0.5 and 0.9\,AU from the star \citep{guedes2008}. Very recently, based on a substantial body of RV data, \citet{dumusque2012} proposed that an Earth-mass planet orbits \acenb\ with a three day period \citep[but see][]{hatzes2013}. In other words, the radial distance of \acenb b (0.04\,AU), which corresponds to only nine stellar radii, is evidently far interior to the HZ, and therefore the surface conditions should be far from being able to support any form of life as we know it.

Based on sensitive {\it Herschel} \citep{pilbratt2010} observations, a relatively large fraction of stars with known planets exhibit detectable far-infrared excess emission due to cool circumstellar dust  \citep{eiroa2011,marshall2013,krivov2013}, akin to the debris found in the asteroid (2--3\,AU) and Edgeworth-Kuiper (30--55\,AU) belts of the solar system. As part of the {\it Herschel} Open Time Key Programme DUNES \citep[DUst around NEarby Stars:][]{eiroa2013} we observed \acen\ to search for dust emission associated with the stars, which is thought to originate on planetesimal size scales and being ground down to detectable grain sizes by mutual collisions. Persistent debris around the stars would be in discs of a few AU in size and would re-emit intercepted starlight in the near- to mid-infrared. To within 16\%, ISO-SWS observations did not detect any excess above the photosphere of \acen\,A between 2.4 and 12 \,\um\ \citep[][and Fig.\,\ref{SED} below]{decin2003}. On the other hand, a circumbinary Edgeworth-Kuiper belt analogue would be much larger and the dust much cooler, so that it would emit predominantly at far-infrared (FIR) and submillimetre (submm) wavelengths. Such a belt would be spatially resolved with the Herschel beam, in principle allowing the detection of structures due to dynamical interactions with e.g. a binary companion and/or giant planets \citep[e.g.,][and references therein]{wyatt2008}, but its surface brightness could be expected to be very low, rendering such observations very difficult. 

\begin{figure}
  \resizebox{\hsize}{!}{
    \rotatebox{00}{\includegraphics{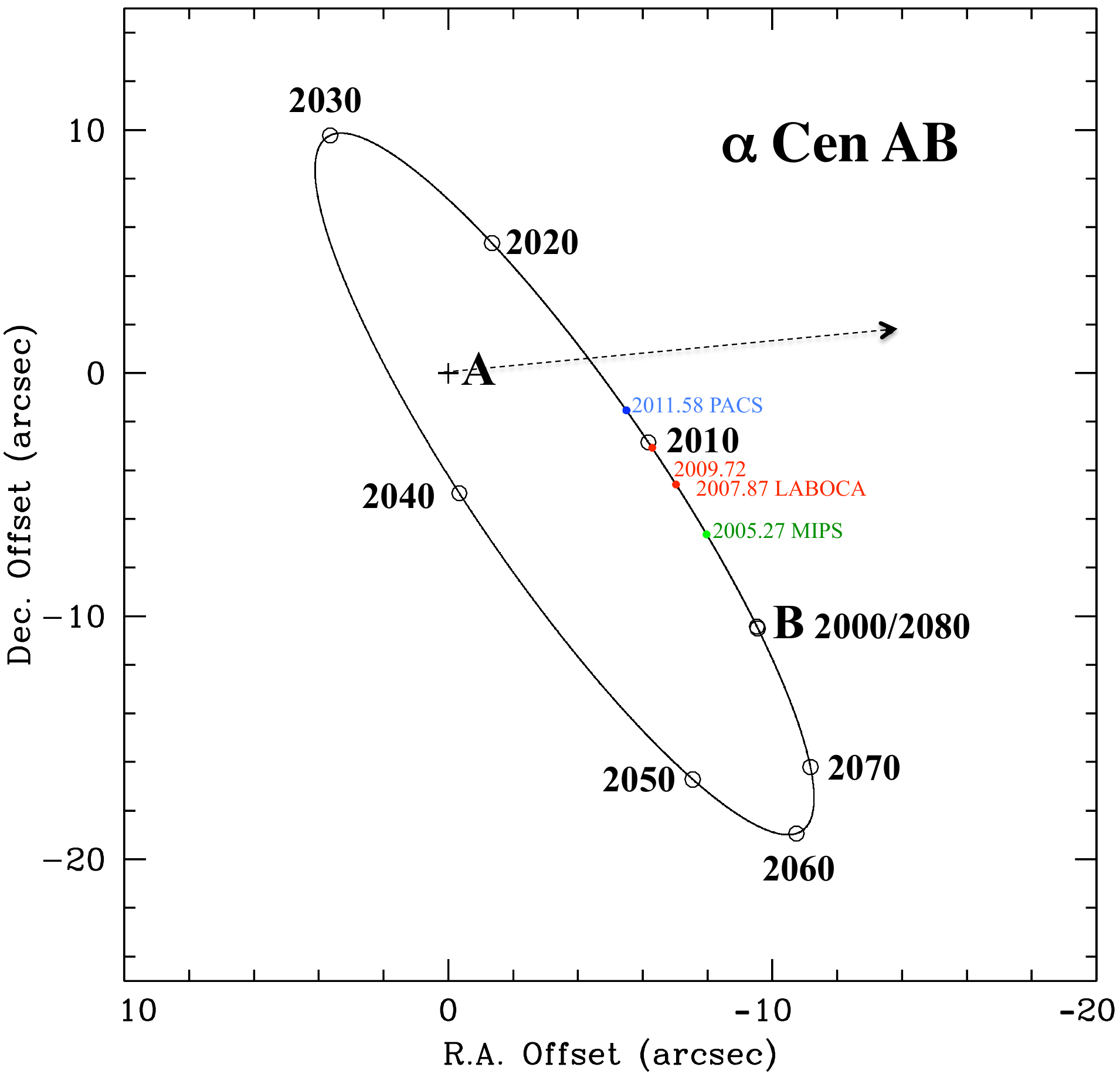}}
                        }
  \caption{The binary \acen\,AB as it appears in the sky at various occasions. Shown is the orbit of \acenb\ with respect to the primary A, being at the origin and with north up and east to the left. During the observing period the separation of the stars was largest in 2005 and smallest in 2011. The relative positions in the PACS (blue) and LABOCA (red) maps are indicated. The proper motion of \acen\ for the time span between the LABOCA and PACS observations is shown by the dashed arrow. In addition, the separation between the stars at the time of the MIPS observations is indicated by the green dot. Orbital elements are adopted from \citet{pourbaixetal2002}.
   }
  \label{orbit}
\end{figure}

The DUNES programme focusses on nearby solar-type stars and the observations with {\it Herschel}-PACS \citep{poglitsch2010} at 100\,\um\ and 160\,\um\ are aimed at detecting the stellar photospheres at an $S/N\ge5$ at 100\,\um\ and has observed 133 nearby Sun-like stars (FGK type, $d < 25$ pc). Prior to {\it Herschel}, little data at long wavelengths which have high photometric quality are available for \acena\ and B. One reason likely being detector saturation issues  due to their brightness (e.g., WISE bands W1--W4), another is due to contamination in the large beams by confusing emission near the galactic plane (e.g., IRAS, ISO-PHOT and AKARI data).

\begin{table*}
    \caption{Observing Log}
    \label{obs_log}
    \begin{tabular}{llcclrlc}      
\hline\hline    
\noalign{\smallskip}             \noalign{\smallskip}	
Instrument/ 		& Obs/Pgm ID	&  Wavelength		 	  & Beam width  	& Observing Date	& $t_{\rm int}$	& Centre Coordinates$^a$ 		& Offset$^b$ 	\\
Mode			& 			& $\lambda_{\rm eff}$ (\um)& HPBW (\asec)	&  year--mo--day	& (sec)		&  h   m   s\hspace{1.0cm}\adeg\   \amin\   \asec    	& (\asec)		 \\
\noalign{\smallskip}	
\hline                        	
\noalign{\smallskip}	
{\it Herschel} PACS	& 1342224848	& 100 		& \phantom{1}\asecdot{7}{7} 	& 2011--07--29		&    542		&14 39 30.200   $-60$ 49 59.66    	& 1.7 	\\
				& 1342224849	& 160		& \asecdot{11}{3}			&  2011--07--29	&    542		&14 39 30.115   $-60$ 49 59.54	& 1.3		\\
SpirePacsParallel	& 1342203280	&   70		& \phantom{1}\asecdot{5}{9}\,$\times$\,\asecdot{12}{2}	&  2010--08--21	& 9490		&14 28 59.831 	  $-60$ 39 01.60	& 2\adeg$ \times$ 2\adeg\ map \\
				& 1342203281	& 160		& \phantom{1}\asecdot{11}{6}\,$\times$\,\asecdot{15}{6}	&  2010--08--21	& $\dots$		& $\dots$						& $\dots$          \\
				& $\dots$		&  250		& \asecdot{17}{6}$^{c}$		&  2010--08--21	& $\dots$		& $\dots$						&  $\dots$        \\	
				& $\dots$		&  350		& \asecdot{23}{9}$^{c}$		&  2010--08--21	& $\dots$		& $\dots$						&  $\dots$          \\	
				& $\dots$		&  500		& \asecdot{35}{2}$^{c}$		&  2010--08--21	& $\dots$		& $\dots$						&  $\dots$          \\				
APEX-LABOCA	&384.C-1025(A)& 870 		& \asecdot{19}{5}			& 2009--09--19		&   7147		& 14 39 31.725   $-60$ 49 59.10	&  3.7     \\
				&380.C-3044(A)& 870		& \asecdot{19}{5} 			& 2007--11--10 to 13& 34026		& 14 39 32.349   $-60$ 50 00.00	&  3.6     \\
APEX-SHeFI		&090.F-9322(A)& 1300		& \asecdot{27}{1} 			& 2012--08--16		&   2646		& 14 39 35.060	   $-60$ 50 15.1	& sp. line map	\\
\noalign{\smallskip}	
\hline
    \end{tabular}
    \begin{list}{}{}
    \item[$^{a}$] Observed equatorial coordinates (J2000) toward the target, i.e. the centre of the fitted point source, which refers to the primary \acena, except for Hi-GAL and APEX-SHeFI maps. 
    \item[$^{b}$] Offset of observed with respect to intended (= commanded) position. Except for SHeFI, coordinates are corrected for the stellar proper and orbital motions. Not applicable (na) to survey data.
    \item[$^{c}$] According to the SPIRE manual, see {\small \texttt{http://herschel.esac.esa.int/Docs/SPIRE/html/}}, the average beam area is 423\,arcsec$^2$ at 250\,\um, 751\,arcsec$^2$ at 350\,\um\ and 1587\,arcsec$^2$ at 500\,\um\ . 
    \end{list}
\end{table*}

Due to their proximity, and having an age comparable to that of the Sun \citep[4.85\,Gyr,][]{thevenin2002}, \acen\ is an excellent astrophysical laboratory for ``normal" low-mass stars, otherwise known to be very difficult to calibrate, not the least with respect to their ages. From numerous literature sources, \citet{torresetal2010} have compiled the currently best available basic stellar parameters of the \acen\ system. The given errors on the physical quantities are generally small. However, whereas the tabulated uncertainty of the effective temperature of e.g. \acena\ is less than half a percent, the observed spread in Table\,1 of \citet{porto2008} corresponds to more than ten times this much. On the other hand, the radii given by \citet{torresetal2010} are those directly measured by \citet{kervella2003} using interferometry, with errors of 0.2\% and 0.5\% for A and B, respectively \citep{bigot2006}. Masses have been obtained from astroseismology and are good to within 0.6\% for both components \citep{thevenin2002}. 

For such an impressive record of accuracy for the stellar parameters of the \acen\ components it should be possible to construct theoretical model photospheres with which observations can be directly compared to a high level of precision. Here, we report PACS observations of \acen\ at 100\,\um\ and 160\,\um. These single-epoch data are complemented by LABOCA \citep{siringo2009} data at 870\,\um\ obtained during two different epochs. The LABOCA observations primarily address two issues: the large proper motion (\asecdot{3}{7}\,yr$^{-1}$) should enable the discrimination against background confusion and, together with SPIRE photometry (see below), these submm data should also provide valuable constraints on the Spectral Energy Distributions (SEDs). This could potentially be useful to quantify some of the properties of the emitting dust and/or to gauge the temperature minima at the base of the stellar chromospheres. A clear understanding of the latter is crucial when attempting to determine extremely low levels of cool circumstellar dust emission. 

The paper is organised as follows: Sect.\,2 outlines our observations with various facilities, both from space and the ground. As we are aiming at low-level detections, the reduction of these data is described in detail. Our primary results are communicated in Sect.\,3. In the discussion section, 4, we examine the lower chromospheres of the \acen\ stars in terms of the radiation temperatures from their FIR-photospheres. Possible contributions to the FIR/submm SEDs by dust are also addressed, using both analytical estimations and detailed numerical models. Finally, Sect.\,5 provides a quick overview of our conclusions.

\section{Observations and data reduction}

\subsection{{\it Herschel}}

In the framework of our observing programme, i.e. the DUNES Open Time Key Programme, PACS photometric images were obtained at 100\,\um\ and 160\,\um. In addition, from the Hi-GAL survey (PI S. Molinari), we acquired archive data for \acen\ at 70\,\um\ and 160\,\um\ obtained with PACS and at 250\,\um, 350\,\um\ and 500\,\um\ with SPIRE \citep{griffin2010}. The relative position of the stars in the sky during the observational period is shown in Fig,\,\ref{orbit}.
 
\subsubsection{DUNES: PACS 100\,\um\ and 160\,\um}

PACS scan maps of  \acen\ were obtained at 100\,\um\ and 160\,\um\ at two array orientations (70\adeg\ and 110\adeg) to suppress detector striping. The selected scan speed was the intermediate setting, i.e. 20\asec s$^{-1}$, determining the PSF at the two wavelengths (\asecdot{7}{7} and 12\asec, respectively). The 100\,\um\ filter spans the  region 85--130\,\um\ and the 160\,\um\ filter 130--210\,\um\ and the observations at 100 and 160\,\um\ are made simultaneously. The data were reduced with HIPE v.8.0.1. The native pixel sizes are \asecdot{3}{2} and \asecdot{6}{4} at 100 and 160\,\um, respectively, and in the reduced images the resampling resulted in square pixels of 1\asec\ at 100\,\um\ and 2\asec\ at 160\,\um. 

The two stellar components are approximated by model instrument PSFs of the appropriate wavelength based on an observation of $\alpha$ Boo rotated to match the telescope position angle of the $\alpha$ Cen observations. It is important to match the orientation of the PSF due to the non-circular tri-lobal structure of the \textit{Herschel} PACS PSF which exists below the 10\% peak flux level.

\begin{table}
    \caption{Photometry and FIR/flux densities of \acen tauri}
    \label{fluxes}
    \begin{tabular}{llcc}      
\hline\hline    
\noalign{\smallskip}             
    $\lambda_{\rm eff}$	&  \acena			& \acenb  			&	Photometry	\\
    (\um)				&$S_{\nu}$ (Jy)	& $S_{\nu}$ (Jy)	&   \& Reference    \\
\noalign{\smallskip}	
\hline
\noalign{\smallskip}	
   0.440  	&  $2215 \pm 41$   	&   $536 \pm 10$  	 &	B (1)           \\
   0.550  	&  $3640 \pm 67$   	&   $1050 \pm 19$ 	 &	V (1)           \\
   0.790  	&  $4814 \pm 89$  	&  $1654 \pm 30$	 &	I  (1)          \\
   0.440  	&  $2356 \pm 43$   	&  $572 \pm  10$	 &	B  (2)          \\
   0.550  	&  $3606 \pm 66$  	&  $1059 \pm 20$	 &	V  (2)          \\
   0.640  	&  $4259 \pm 78$   	&  $1387 \pm 26$	 &	R$_{\rm c}$ (2)	\\
   0.790  	&  $4784 \pm 88$   	&  $1666 \pm 31$	 &	I $\!_{\rm c}$ (2) \\
   1.215  	&  $4658 \pm 86$   	& $1645 \pm 30$	 &	J   (3)             \\
   1.654  	&  $3744 \pm 69$   	& $1649 \pm 31$	 &	H  (3)           	\\
   2.179  	&  $2561 \pm 47$   	& $1139 \pm 21$	 &	K  (3)            	\\
   3.547  	&  $1194 \pm 22$   	& $521 \pm 10$ 	 &	L  (3)            	\\
   4.769  	&  $592  \pm 11$   	& $258 \pm 5$		 &	M  (3)          \\
  24        & $30.84 \pm 0.76$	& $13.63 \pm 0.33$	 &	MIPS (4)   	 	\\
  70       	&  $3.35 \pm 0.28$ & $1.49 \pm 0.28$	 &	PACS (5) 	    \\
100      	&  $1.41 \pm 0.05$   & $0.67 \pm 0.037$	 &	PACS (6)	    \\
160      	&  $0.56 \pm 0.06$   & $0.21 \pm 0.06$	 &$^{\ast}$	PACS  (5), (6) \\
250	    	&  $0.24  \pm 0.05$  & $0.11 \pm 0.05$      &$^{\ast}$SPIRE (5)	\\
350      	&  $0.145 \pm 0.028$&$0.064 \pm 0.028$ &$^{\ast}$SPIRE (5)	\\
500		&   $0.08 \pm 0.03$   & $0.04 \pm 0.03$     &$^{\ast}$SPIRE (5)	\\
870      	&  $0.028\pm 0.007$& $0.012 \pm 0.007$ &$^{\ast}$LABOCA (7) \\
\noalign{\smallskip}	
\hline
\noalign{\smallskip}	
    \end{tabular}

    \begin{list}{}{}
    \item[$^{\ast}$] Asterisks indicate values determined using $S_{\nu,\,{\rm A}}/S_{\nu,\,{\rm B}}  = 2.25$ \citep[see ][]{liseau2013}.
    \item[(1)] HIPPARCOS,  (2) \citet{bessell1990}, (3) \citet{engels1981}.
    \item[(4)] A.\,Mora [priv. com.; FWHM(24\,\um) = 6\asec]. Binary separation on 9 April, 2005, \asecdot{10}{4}
    \item[(5)] Hi-GAL: KPOT\_smolinar\_1, fields 314\_0 \& 316\_0. {\it Herschel}-beams FWHM(70\,\um) = \asecdot{5}{6}, (100\,\um) = \asecdot{6}{8}, (160\,\um) = \asecdot{11}{3}, (250\,\um) = \asecdot{17}{6}, (350\,\um) = \asecdot{23}{9}, (500\,\um) = \asecdot{35}{2}. Binary separation on 21 August, 2010, \asecdot{6}{3}. 
    \item[(6)] DUNES: KPOT\_ceiroa\_1. Binary separation 29 July, 2011, \asecdot{5}{7}. 
    \item[(7)] 384.C-1025, 380.C-3044(A): FWHM(870\,\um) = \asecdot{19}{5}. Binary separation 20-13 November, 2007, \asecdot{8}{8} and 19 September, 2009, \asecdot{7}{0}.
    \end{list}
\end{table}

Fitting of the two components was carried out by subtraction of the components in series, starting with the brighter. Each component model PSF was scaled to the estimated peak brightness and shifted to the required positional offset from the observed source peak before subtraction. Note that the position offset of B relative to A was fixed.

As described in detail by \citet{eiroa2013} the level of the background and the rms sky noise were estimated by calculating the mean and standard deviation of 25 boxes sized 9\arcsec$\times$9\arcsec at 100~$\mu$m and 14\arcsec$\times$14\arcsec at 160~$\mu$m scattered randomly at positions lying between 30\arcsec-- 60\arcsec\ from the image centre and within the area of which no pixel was brighter than twice the standard deviation of all non-zero pixels in the image (being the threshold criterion for source/non-source determination while high pass filtering during map creation). The calibration uncertainty was assumed to be 5\% for both 100~$\mu$m and 160~$\mu$m \citep{balog2013}.

The PACS calibration scheme is further described in detail in {\small \texttt{http://herschel.esac.esa.int/twiki/bin/view/Public/\\
PacsCalibrationWeb\#PACS\_instrument\_and\_calibration}}. Aperture (and potential colour) corrections of the stellar flux densities and sky noise corrections for correlated noise in the super-sampled images are described in the technical notes PICC-ME-TN-037 and PICC-ME-TN-038 and {\small \texttt{https://nhscsci.ipac.caltech.edu/sc/index.php/Pacs/Ab-\\
soluteCalibration}}.

\begin{figure*}
  \resizebox{\hsize}{!}{
    \rotatebox{00}{\includegraphics{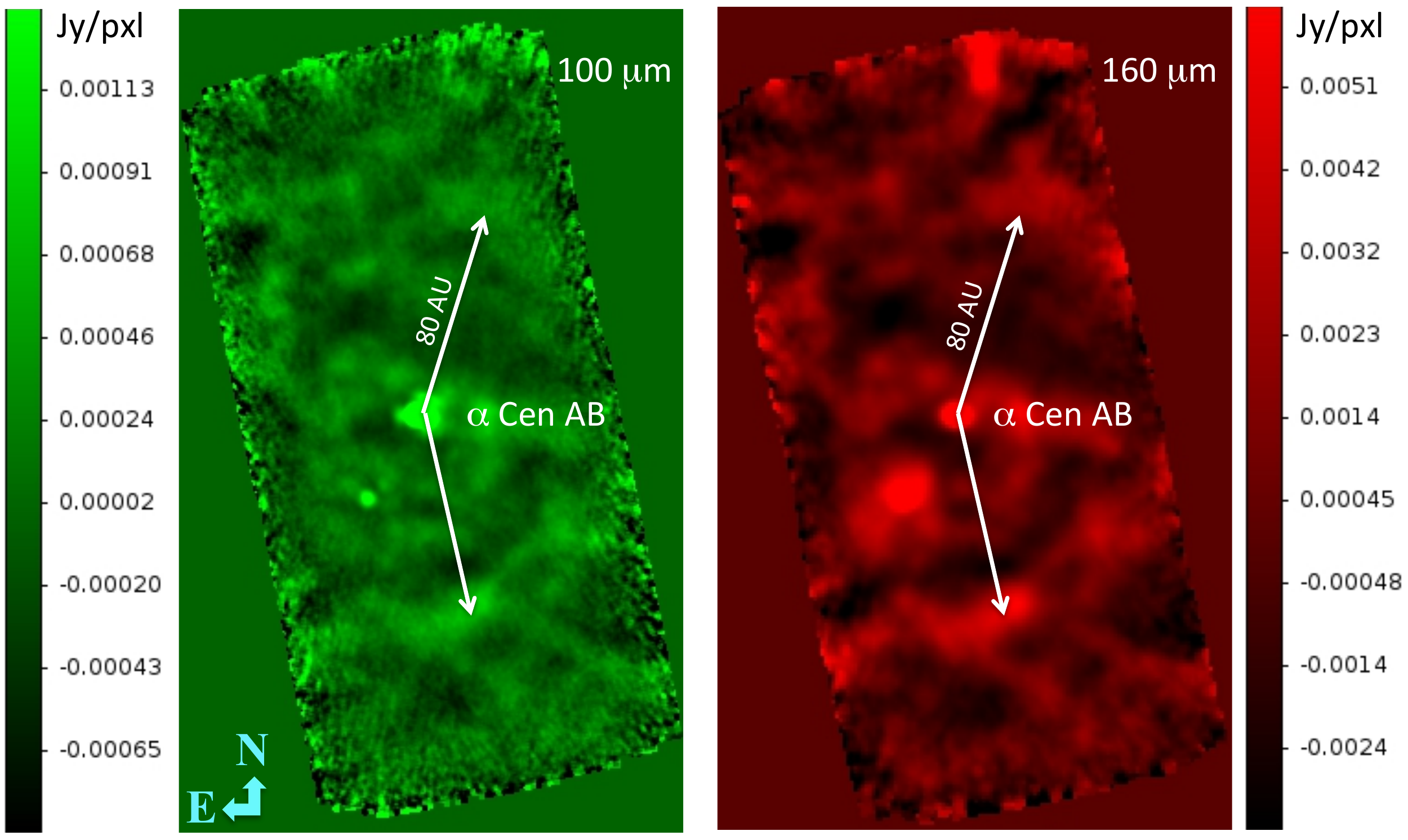}}
                        }
  \caption{The binary \acen\ at 100\,\um\ (left: {\it green}) and 160\,\um\ (right: {\it red}) at high contrast. Scales are in Jy/pxl, where square pixels are 1\asec\ on a side at 100\,\um\ and 2\asec\ on a side at 160\,\um. The field of view is \amindot{1}{75}\,$\times $\,\amindot{3}{5}. The image of \acen\ is extended in the direction of the companion and, at 100\,\um, the pair is quasi-resolved, having a separation of \asecdot{5}{7} along position angle 254\adeg. The white arrows of length 80\,AU correspond approximately to the stable circumbinary regime around \acen\,AB. These point to parts of  coherent structures that are also are seen at all FIR/submm wavelengths. 
     }
  \label{observations}
\end{figure*}

\begin{figure}
  \begin{center}
  \includegraphics[width=88mm]{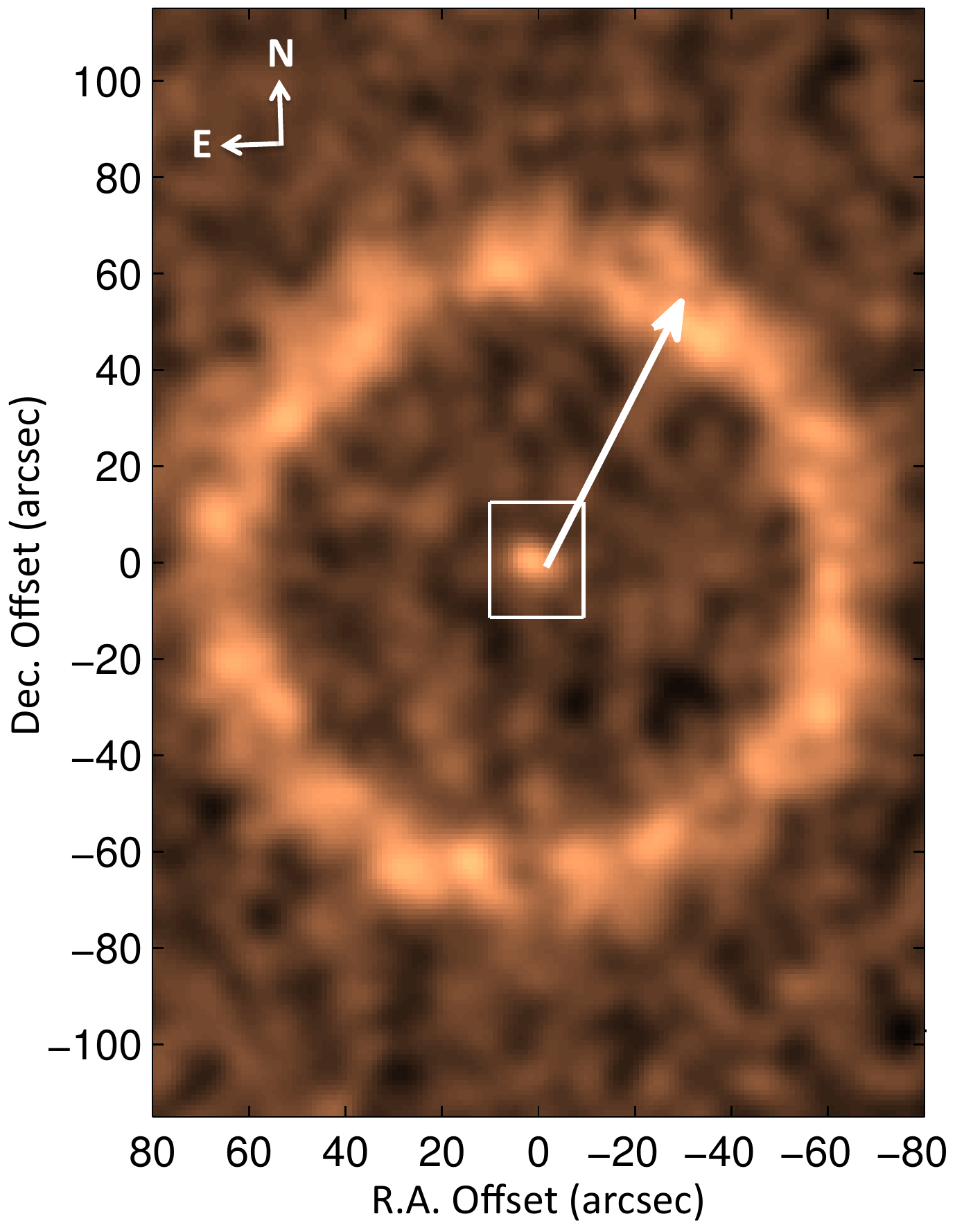}
    \caption{Simulated observation at 160\,$\mu$m of \acena B, assuming a face-on circumbinary dust disc/ring ($m_{\rm d} \sim 0.1\,M_{\leftmoon}$, cf. Sect.\,4.5.2). As seen from the system's barycenter (white rectangle), the ring extends from $70$ to $105$\,AU (the white arrow has a length of 80\,AU). In this image, each pixel is $1 \farcs 04$ on a side. The image was smoothed with a $16 \times 16$ pixel Gaussian filter with a standard deviation of 3 pixels. The added noise is also Gaussian distributed and smoothed to mimick  the noise and beam size of Fig.\,\ref{observations}.}
	\label{particleskyprojection}
  \end{center}
\end{figure}

\subsubsection{Hi-GAL: PACS 70\,\um\ and 160\,\um}

Being part of the Hi-GAL program, these data were obtained at a different scan speed, i.e. the fast mode at 60\asec s$^{-1}$. The field $314\_0$, containing \acen tauri, was observed at both wavelengths simultaneously and in parallel with the SPIRE instrument (see next section). The scanned area subtends 2\adeg $\times$ 2\adeg\ and these archive data are reduced to {\it Level 2.5}. Compared to the data at  longer wavelengths, background problems are much less severe at 70\,\um. The PACS 70\,\um\ data provide the highest angular resolution of all data presented here (Table\,\ref{obs_log}). As before, no further colour correction was required, but the proper aperture correction (1.22)  was applied. After re-binning, the pixel size is the same as that at 100\,\um, viz. one square arcsecond, whereas the 160\,\um\ pixels are as before two arcseconds squared.

\subsubsection{Hi-GAL: SPIRE 250\,\um, 350\,\um\ and 500\,\um}

\begin{figure*}
  \resizebox{\hsize}{!}{
    \rotatebox{00}{\includegraphics{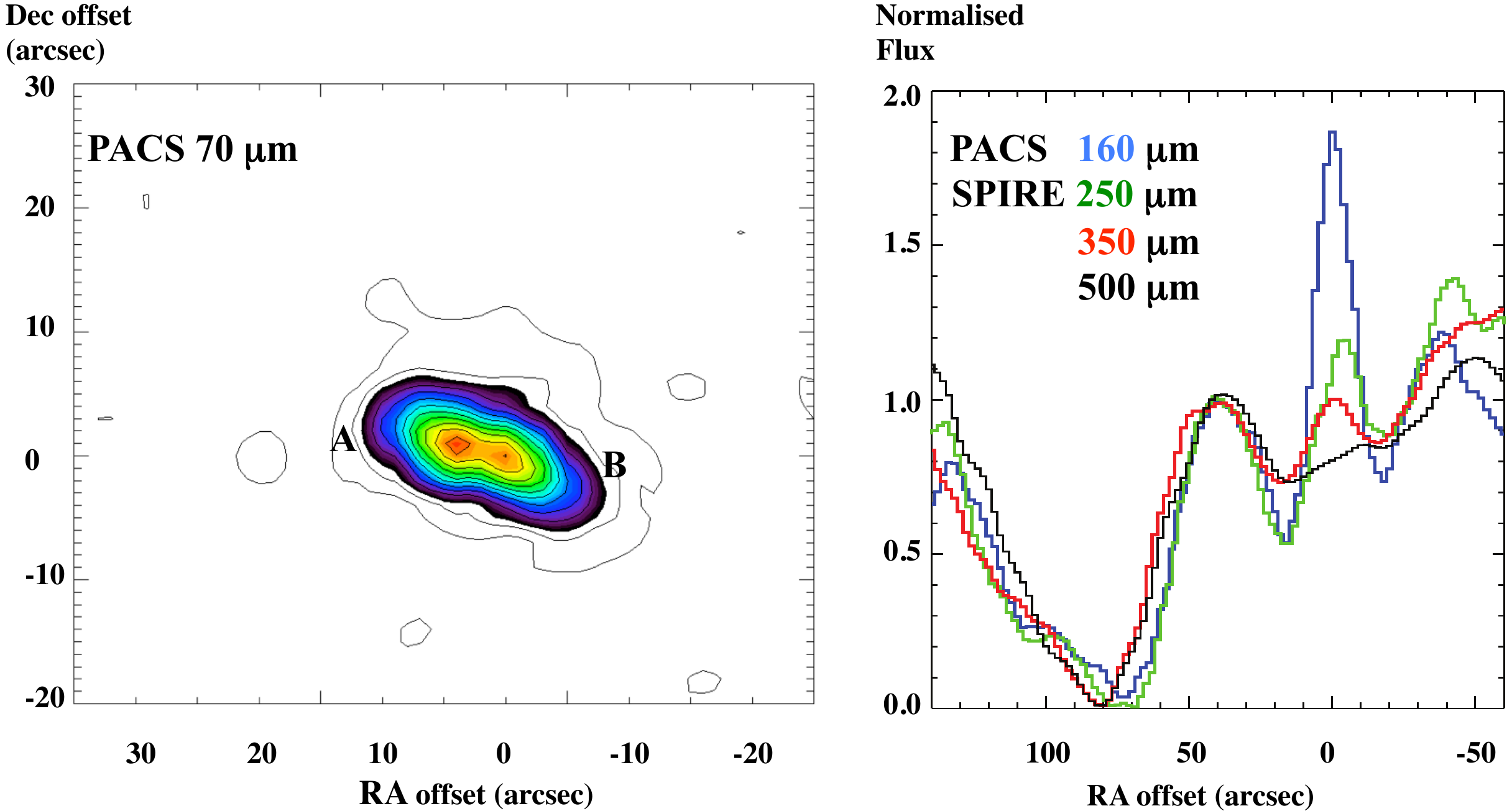}}
                        }
  \caption{{\bf Left:} At 70\,\um\ with PACS, the binary \acena B is clearly resolved (Hi-GAL data).  Note that, as a result of the re-binning using cubic spline interpolation, the stars appear too close in this image. {\bf Right:} The method for the estimation of the source and background fluxes in the SPIRE images is illustrated. Shown is a cut in Right Ascension through the image, normalised to the interstellar dust feature 40\asec\ east of \acen. For comparison, also the 160\,\um\ PACS data are shown in blue. For SPIRE, green identifies 250\,\um, red 350\,\um\ and black 500\,\um. 
   }
  \label{SPIRE}
\end{figure*}

At the relatively low resolution of the SPIRE observations, the strong and varying background close to \acen\ presented a considerable challenge for the flux measurements. However, for wavelengths beyond 160\,\um, the dust emission from the galactic background is expected to be in the Rayleigh-Jeans (RJ) regime and the emissivity should remain reasonably constant. Therefore, when normalised to a specific feature, the background is expected to look essentially the same at all wavelengths, disregarding the slight deterioration of the resolution due to the smearing at the longest wavelengths. The result of this procedure is shown in Fig.\,\ref{SPIRE}, where \acen\ is represented by the scaled PSFs and the background-subtracted fluxes are reported in Table\,\ref{fluxes}. Assuming 10--30\% higher (lower) fluxes for \acen\ results in depressions (excesses) at the stellar position, which are judged unrealistic. This should therefore provide a reasonable estimate of the accuracy of these measurements. The photometric calibration of SPIRE is described by \citet{bendo2013}.

\subsection{APEX}

The Atacama Pathfinder EXperiment (APEX) is a 12\,m submillimetre (submm) telescope located at 5105\,m altitude on the Llano de Chajnantor in Chile. According to the APEX home page\footnote{ \texttt{http://www.apex-telescope.org/telescope/}}, the telescope pointing accuracy is 2\asec\ (rms). The general user facilities include four heterodyne receivers within the approximate frequency bands 200-1400\,GHz and two bolometer arrays, centered at 345\,GHz (870\,\um) and 850\,GHz (350\,\um), respectively. We used LABOCA and SHeFI APEX-1 for the observations of \acen tauri.

\subsubsection{LABOCA 870\,\um}

The Large APEX BOlometer CAmera (LABOCA) is a submillimetre array mounted at the APEX telescope \citep{siringo2009}. The operating wavelength is 870\,\um\ centered on a 150\,\um\ wide window (345 and 60\,GHz, respectively).   With separations of 36\asec, the 295 bolometers yield a circular field of view of \amindot{11}{4}. The angular resolution is HPBW=\asecdot{19}{5} and the under-sampled array is filled during the observations with a spiral mapping method. 

The mapping observations of \acen\  were made during two runs, viz. on November 10--13, 2007, and on September 19, 2009. The data associated with the programmes 380.C-3044(A) and 384.C-1025(A) were retrieved from the ESO archive. As this observing mode is rather inefficient for point sources, we performed additional test observations with the newly installed telescope wobbler on May 20 and July 13--14, 2011. While centering \acenb\ on LABOCA channel 71, the chopping was done by a fixed $\pm 25$\asec\ in east-west direction, which resulted in asymmetric sky-flux pickup in this confused field and these data had therefore to be discarded.

The map data were reduced and calibrated using the software package CRUSH\,2 developed by Attila\,Kov\' acs, see {\small \texttt{http://www.submm.caltech.edu/$\sim$sharc/crush/download.htm}}. The data have been smoothed with a Gaussian of HPBW=13\asec, resulting in an effective FWHM=\asecdot{23}{4}. However, fluxes in Jy/beam are given for an FWHM=\asecdot{19}{5}.

\subsubsection{SHeFI 230\,GHz (1300\,\um)}

Complementing observations were made with the Swedish Heterodyne Facility Instrument (SHeFI) APEX-1 during Director's Discretionary Time (DDT) on August 16, 2012. APEX-1 is a single-sideband receiver and operates in the 230\,GHz band (211--275\,GHz) with an IF range of 4--8\,GHz. At the CO\,(2--1) frequency, 230.5380\,GHz, the half-power-beam-width is 27\asec. The estimated main beam efficiency is $\eta_{\rm mb}=0.75$ and the Kelvin-to-Jansky conversion for the antenna is 39\,Jy/K at this frequency. 

The observing mode was on-the-fly (OTF) mapping with a scan speed of 6\asec\,s$^{-1}$, which resulted in a data cube for the 5\amin\,$\times $\,5\amin\ spectral line map with 441 read-out points (Fig.\,\ref{co_map}). The reference, assumed free of CO emission, was at offset position ($+1500$\asec, $-7200$\asec). During the observations, the system noise temperature was typically \tsys\,=\,200\,K. As backend we used the Fast Fourier Transform Spectrometer (FFTS) configured to 8192 channels having velocity resolution 0.16\,\kms, yielding a total Doppler bandwidth of 1300\,\kms. 

These ON-OFF observations generate a data cube, with two spatial dimensions (the ``map") and one spectral dimension (intensity vs frequency spectrum). The data were reduced with the xs-package by  P.\,Bergman (\texttt{http://www.chalmers.se/rss/oso-en/observations/ data-reduction-software}). This included standard, low order, base line fitting and subtraction, yielding for each ``pixel"  a spectrum with the intensity given relative to the zero-Kelvin level (see Fig.\,\ref{co_map}\,a).

\section{Results}

\subsection{{\it Herschel} PACS and SPIRE}

The stars were clearly detected at both PACS 100\,\um\ and 160\,\um, as was expected on the basis of the adopted observing philosophy of the DUNES programme. The last entry in Table\,\ref{obs_log} refers to the difference between the proper and orbital motion corrected commanded and the observed position determined from the fitting of Gaussians to the data, and these are found to be clearly less than 2\asec. This compares very favourably with the findings for PACS of the average offset of  \asecdot{2}{4} by \citet{eiroa2013} for  a sample of more than 100 solar type stars. Similarly, also for the LABOCA data, observed offsets are acceptable, i.e. within $2 \sigma$ of the claimed pointing accuracy (Table\,\ref{obs_log}).

Observed flux densities of \acena\ and B are given in Table\,\ref{fluxes}, where the quoted errors are statistical and refer to relative measurement accuracy only. Absolute calibration uncertainties are provided by the respective instrument teams and cited here in the text. Together with complementing {\it Spitzer} MIPS data, these data are displayed in Figs.\,\ref{SED}. For both \acena\ and B a marginal excess at 24\,\um\ corresponding to 2.5\,$\sigma$ and 2.6\,$\sigma$ is determined respectively. The measured flux ratio at this wavelength corresponds well with the model ratio, i.e. $2.26 \pm 0.11$ and 2.25 respectively.

\acen\ was detected at all SPIRE wavelengths, but with a marginal result at 500\,\um\ (Fig.\,\ref{SPIRE_LABOCA} and Table\,\ref{fluxes}). As is also evident from these images, the pair was not resolved.

\subsection{APEX LABOCA}

\begin{figure}
  \resizebox{\hsize}{!}{
    \rotatebox{00}{\includegraphics{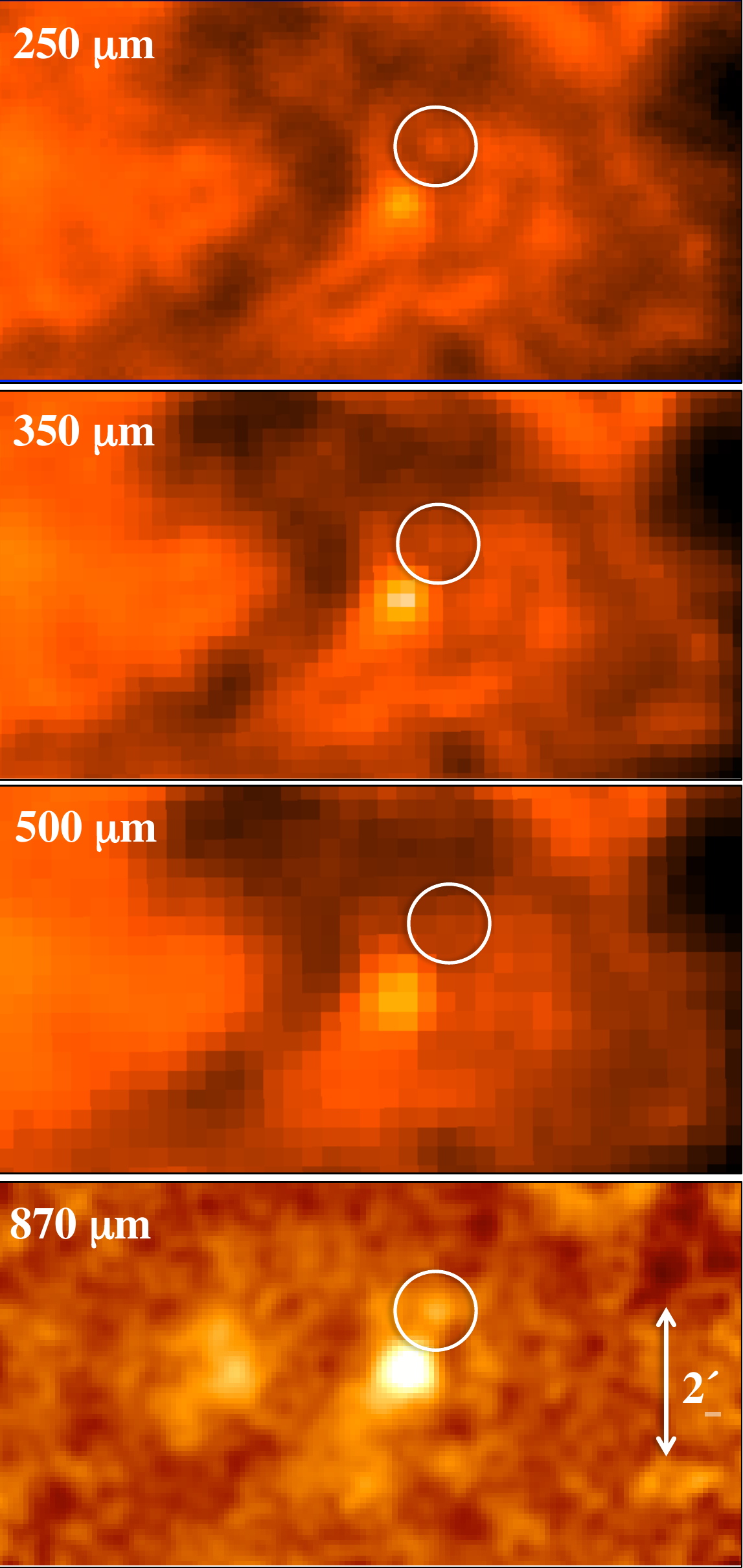}}
                        }
  \caption{Gallery of SPIRE and LABOCA data for \acen tauri which is inside the circles. Top to bottom: 250\,\um, 350\,\um\ and 500\,\um\ (SPIRE) and 870\,\um\ (LABOCA), where the length of the arrows corresponds to 2\amin. }
  \label{SPIRE_LABOCA}
\end{figure}

The LABOCA fields are significantly larger than those of the PACS frames and contain a number of mostly extended sources of low intensity. However, \acen\ is clearly detected (Fig.\,\ref{SPIRE_LABOCA}), but the binary components are not resolved. In Table\,\ref{fluxes}, only the average is given for the 2007 and 2009 observations, as these flux densities are the same within the errors. 

The proximity of the \acen\ system leads to angular size scales rarely ever encountered among debris discs, which are generally much farther away. The PACS field of view is 105\asec\,$\times$\,210\asec, so that a Edgeworth-Kuiper belt analogue would easily fill the images. The LABOCA frames contain structures similar to those discernable in Fig.\,\ref{observations}. Although at faint levels, most of these sources are definitely real, as they repeatedly show up in independent data sets at different wavelengths, obtained with different instruments and at different times. Knotty, but seemingly coherent, arcs and ring-like features on arcmin scales mimic the morphology of belt features. If these features were associated with the \acen\ stars, they should move in concert with them at relatively high speed. At 870\,\um, the strongest feature lies less than one arcminute southeast of \acen\ and is barely detectable at 100\,\um, but clearly revealed at 160\,\um. Measurements of the sky positions for both this feature (``Bright Spot") and \acen\ are reported in Table\,\ref{laboca_pos}. For the Bright Spot these differ by $+$\asecdot{1}{2} in Right Ascension (R.A.) and $+$\asecdot{0}{9} in Declination (Dec.), which provides an estimate of the measurement error, i.e. \about\,1\asec\ for relatively bright and centrally condensed sources. 

Due to its high proper motion,  for \acen\ these values are significantly larger,  viz. $\Delta\,{\rm R.A.}  = -$\asecdot{6}{8} and $\Delta\,{\rm Dec.}  = +$\asecdot{2}{9}, at position angle PA \about\  247\adeg. Neglecting the \asecdot{0}{1} parallactic contribution, SIMBAD\footnote{\small \texttt{http://simbad.u-strasbg.fr/simbad/}} data for \acen\ yield for the 1.856\,yr observing period $\Delta\,{\rm R.A.}  = -$\asecdot{6}{7}, $\Delta\,{\rm Dec.}  = +$\asecdot{1}{3} and PA = 259\adeg. 

Similar results are obtained involving other features in the 870\,\um\ images. This provides firm evidence that the inhomogenous background is stationary and unlikely part of any circumbinary material around \acen AB. A few notes concerning the nature of this background are given in the next section. 

\begin{table}
\caption{Dual epoch  position data with LABOCA}             
\label{laboca_pos}      
\begin{tabular}{llcc}      
\hline\hline    
\noalign{\smallskip}             
Object			& Year		&  R.A. (J2000)	& Dec. (J2000)	 	\\
\noalign{\smallskip}	
\hline                        
\noalign{\smallskip}				
\acen tauri AB		& 2007		& 14 39 32.174 & $-60$ 50 02.5     \\
				& 2009		& 14 39 31.246 & $-60$ 49 59.6     \\
\noalign{\smallskip}	
{\it Bright Spot}		& 2007		& 14 39 33.953 & $-60$ 50 44.5     \\
				& 2009		& 14 39 34.120 & $-60$ 50 43.6     \\				
\hline                                   
\end{tabular}
\end{table}

\begin{figure}
  \resizebox{\hsize}{!}{
    \rotatebox{00}{\includegraphics{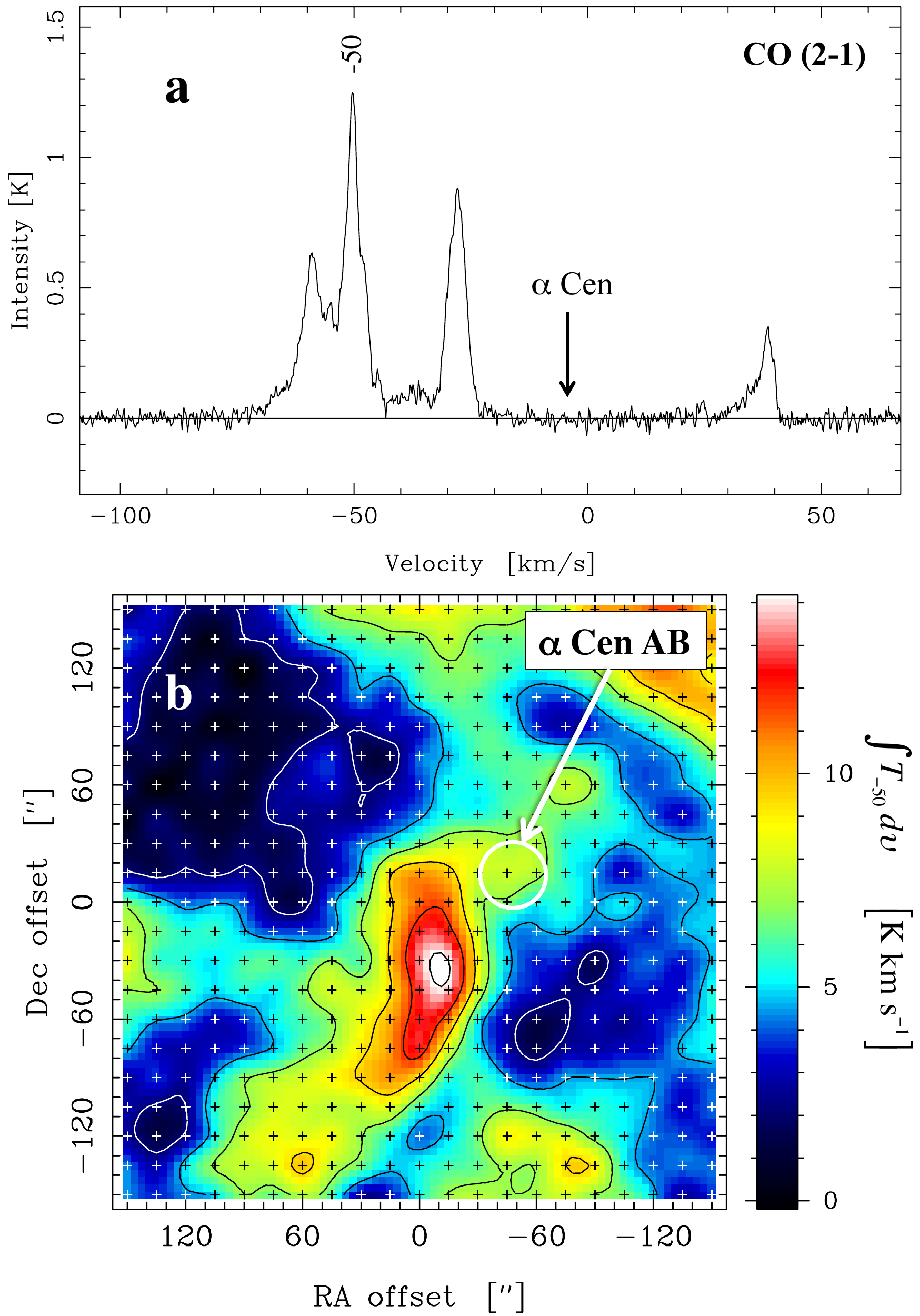}}
                        }
  \caption{{\bf (a)}  Grand average of all CO\,(2--1) spectra, in \tmb\ vs \vlsr, of the 5\amin\,$\times$\,5\amin\ map toward \acen tauri. Strong emission features at several lsr-velocities are evident in this direction of the Galaxy (\ltwo\,=\,\adegdot{315}{73}, \btwo\,=\,\adegdot{$-0$}{68}), with the strongest line at \vlsr\,$=-50$\,\kms.  {\bf (b)}  Integrated intensity map of the CO\,(2--1) line at \vlsr\,=\,$-50$\,\kms, i.e. $\int\!\!T_{-50}\,d\upsilon$, in the direction of \acen tauri [for APEX and at $\nu=230$\,GHz, $\int\!T\,d\upsilon$ (K\,\kms) = $6 \times 10^{15\,}\int\!F_{\!\nu}\,d\nu$ (\ecs)]. The origin of the map is at the J2000 equatorial coordinates of \acen\,B, i.e R.A. = \radot{14}{39}{35}{06} and Dec = \decdot{$-60$}{50}{15}{1}, and read-out positions are marked by crosses. At the time of observation (16 August 2012), the star is located at the centre of the white ring. The bright feature south of the map centre is also prominent in the continuum observations longward of 100\,\um\ (cf. Fig.\,\ref{observations}) and is the dominating source at 870\,\um\ (Fig.\,\ref{SPIRE_LABOCA}).
   }
  \label{co_map}
\end{figure}

\subsection{APEX-1 SHeFI}

In the direction of \acen, strong CO\,(2--1) emission lines are found at Local-Standard-of-Rest (LSR) velocities of $-30$, $-50$ and $-60$\,\kms, with a weaker component also at $+40$\,\kms. This is entirely in accord with the observations of the Milky Way in CO\,(1--0) by \citet{dame2001}. As an example, the distribution of the integrated line intensity of the $-50$\,\kms\ component is shown in Fig.\,\ref{co_map}, providing an overview of the molecular background in this part of the sky. With line widths typical of giant molecular clouds, this confusing emission is certainly galactic in origin and not due to an anonymous IR-galaxy, as the lines would be too narrow. We can also exclude the possibility of a hypothetical circumbinary dust disc around the pair \acena B, since the observed lines fall at unexpected LSR-velocities ($\upsilon_{\alpha\,{\rm Cen}} \sim -3$\,\kms) and also are too wide and too strong. Since these data do not reveal the continuum, the stars \acena B are not seen in the CO line maps.

\section{Discussion}

\subsection{FIR and submm backgrounds}

At a distance of 1.3\,pc and located far below the ecliptic plane, any {\it foreground} confusion can be safely excluded. At greater distances, bright cirrus, dust emission from galactic molecular clouds, PDRs, H\,II regions and/or background galaxies potentially contribute to source confusion at FIR and submm wavelengths. However, the projection of \acen\ close to the galactic plane makes background issues rather tricky (see Sect.\,2.1.3). In particular extragalactic observers tend to avoid these regions and reliable catalogues for IR galaxies are generally not available. However, as our spectral line maps show (Fig.\,\ref{co_map}), the observed patchy FIR/submm background is clearly dominated by galactic emission.

\subsection{The SEDs of \acena\ and \acenb}

\subsubsection{The stellar models}

\begin{figure*}
  \resizebox{\hsize}{!}{
    \rotatebox{00}{\includegraphics{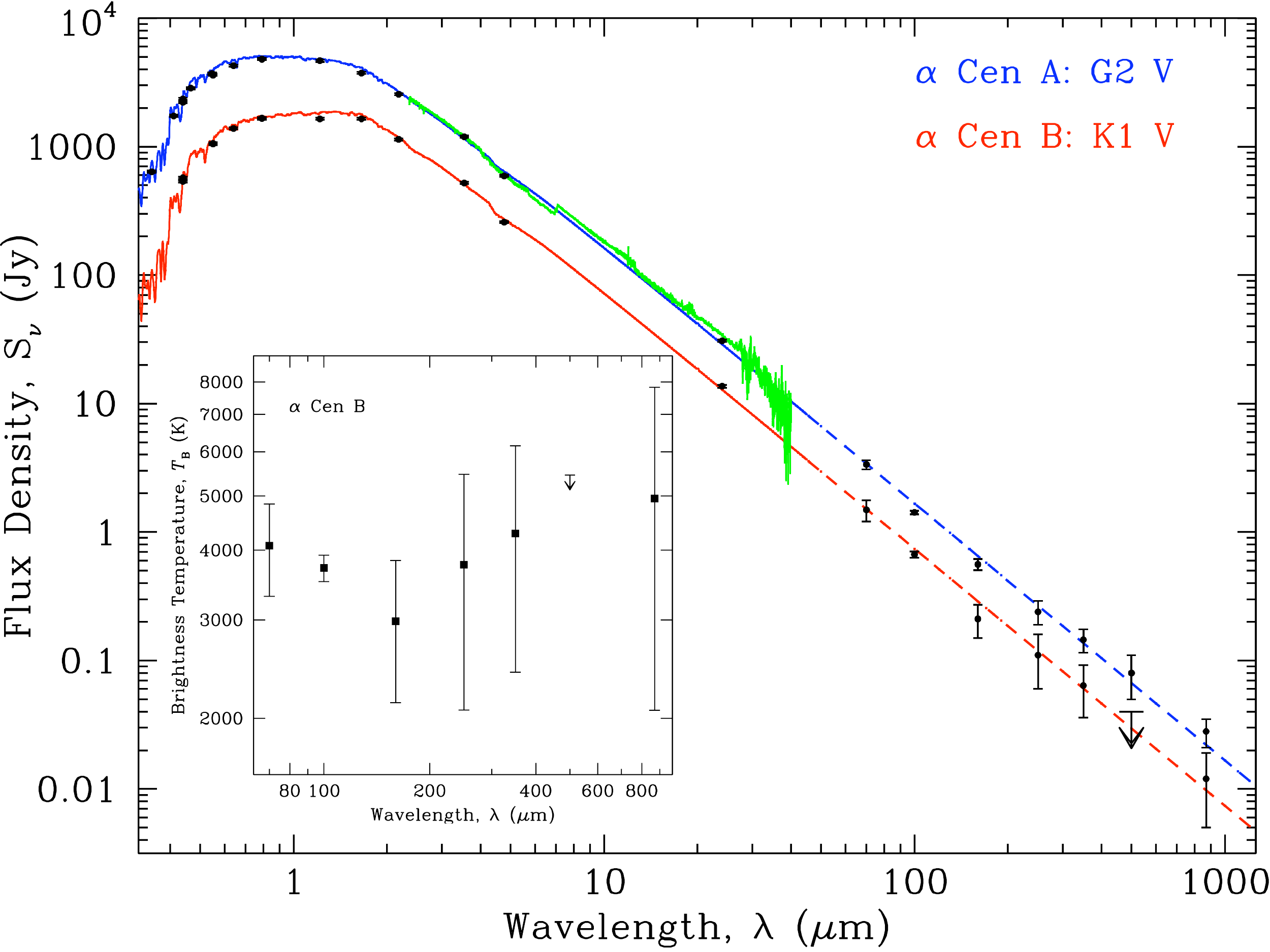}}
                        }
  \caption{SEDs of the binary \acena B. The model photospheres for the individual stars are shown by the blue and red lines, respectively. The PHOENIX model computations extend to about 45\,\um\ (solid lines). At longer wavelengths, these correspond to extrapolations using the proper $B_{\nu}(T_{\rm eff})$ (dashes).  Photometric data points are shown for comparison, with the upper limit being $3 \sigma$ (Table\,\ref{fluxes}). The green curve shows an ISO-SWS low-resolution observation, viz. TDT\,60702006  \citep[PI C. Waelkens; see also][]{decin2003}. The inset displays the SED of the secondary \acenb\ in the FIR/submm spectral region.
   }
  \label{SED}
\end{figure*}

For the interpretation of the observational data we need to exploit reliable stellar atmosphere models and stellar physical parameters. The model photosphere parameters are based on the weighted average as the result of an extensive literature survey. The models have been computed by a 3D interpolation in a smoothed version of the high-resolution PHOENIX/GAIA grid \citep{brott2005} and with the following parameters from \citet{torresetal2010} and the metallicities from  \citet{thevenin2002}: ($T_{\rm eff},\,\log g,\, {\rm [Fe/H]}$)=(5824\,K, 4.3059, +0.195) for \acena\ and (5223\,K, 4.5364, +0.231) for \acenb. These models are shown in Fig.\,\ref{SED}, together with the photometry (Table\,\ref{fluxes}). We wish to point out again that these data have not been used in the analysis, but merely serve to illustrate the goodness of the models. 

As a sanity check we compare the integrated model fluxes with the definition of the effective temperature, $\int\!S_{\nu}\,d\nu/\!\int\!B_{\nu}(T_{\rm eff})\,d\nu$. The luminosities are conserved to within 0.6\% for \acena\ and 0.1\% for \acenb, which is well within the {\it observational} errors of 2\% and 5\%, respectively \citep{torresetal2010}, which in turn are comparable to the errors of the theoretical model \citep[e.g.,][]{gustafsson2008,edvardsson2008}. These small differences can probably be traced back to the re-gridding of the high resolution models onto a somewhat sparser spectral grid.

\subsubsection{The temperature minima of \acena\ and \acenb}

At heights of some five hundred kilometers above the visible photosphere ($h$\,\lapprox\,\powten{-3}\,\rsun), the Sun exhibits a temperature minimum, i.e. $T/T_{\rm eff} < 1$, beyond which temperatures rise into the chromosphere and corona. Theories attempting to explain the physics of the heating of these outer atmospheric layers generally invoke magnetic fields \citep{carlsson1995}, but the details are far from understood and constitute an active field of solar research \citep[e.g., ][]{dalacruz2013}.  Such atmospheric structure can be expected to be quite common also on other, and in particular solar-type, stars. There, the dominating opacity, viz. H$^-$ free-free, limits the visibility to the far infrared photosphere. For instance, on the Sun, $T_{\rm min}$ occurs at wavelengths around 150\,\um. This is very close to the spectral region, where such phenomenon recently has been directly measured for the first time also on another star  \citep[viz. \acen\,A, ][]{liseau2013}, where, in contrast to the Sun, the spatial averaging over the unresolved stellar disc is made directly by the observations. Being so similar in character, \acena\ may serve as a proxy for the {\it Sun as a star} \citep[see also][]{pagano2004}.

This $T_{\rm min}$-effect has also been statistically observed for a larger sample of solar-type stars \citep[see Fig.\,6 of ][]{eiroa2013}, a fact that could potentially contribute to enhancing the relationship between solar physics and the physics of other stars. In the end, such observations may help to resolve a long standing puzzle in solar physics, concerning the very existence of an actual {\it gas temperature} inversion at 500\,km height and whether this is occurring in the Quiet Sun or in Active Regions on the Sun \citep[e.g., ][]{leenarts2011,beck2013}. Theoretical models have hitherto been inconclusive in this regard.

\begin{figure}
  \begin{center}
    \includegraphics[width=88mm]{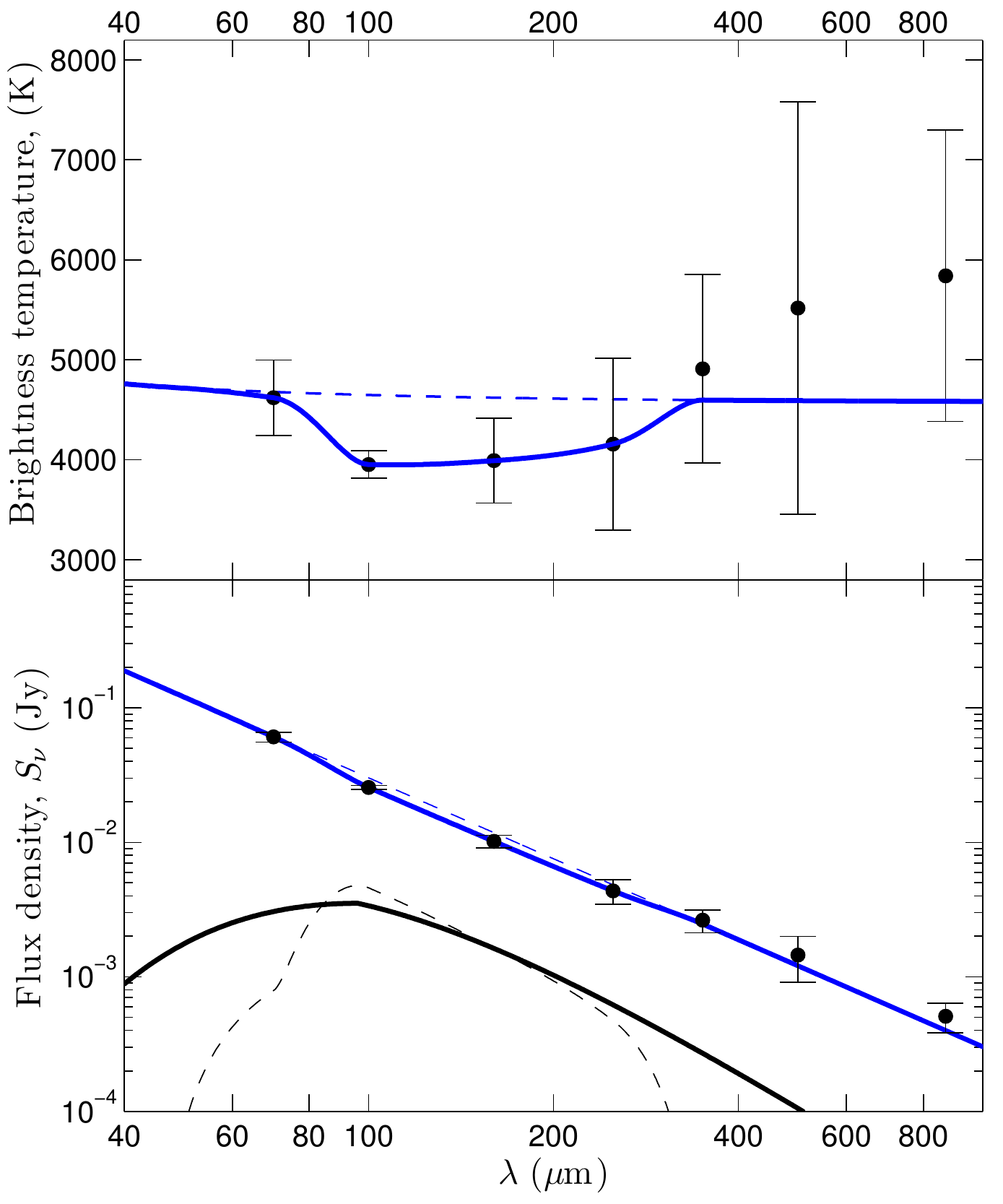}
    \caption{{\bf Top}: The run of the brightness temperature of \acena, where the solid blue line is a fit through the data points around the temperature minimum and the dashed blue line is the stellar blackbody extrapolation. The excess beyond 300\,\um\ is presumed to originate in higher chromospheric layers \citep[e.g., ][]{delaluz2013}. {\bf Bottom}: SED of \acena\ scaled to the distance of 10\,pc. The dashed black curve is the difference between the black body extrapolation (dashed blue line) and the $T_{\rm min}$-fit (solid blue line). The solid black curve is the emission of the fitted 53\,K dust ring.}
	\label{acenatminima}
  \end{center}
\end{figure}

For \acena, \citet{liseau2013} found that $T_{\rm min} = 3920 \pm 375$\,K, corresponding to the ratio $T_{\rm min}/T_{\rm eff} = 0.67 \pm 0.06$. This is lower than that observed for the Sun from spectral lines in the optical, viz. \about\,0.78 from analysis of the Ca \textsc{ii} K-line \citep{ayresetal1976}. These authors also estimated this ratio for \acenb, i.e. 0.71 and 0.72 using $T_{\rm eff} = 5300$ and 5150\,K, respectively. For the data displayed in Fig.\,\ref{SED}, we arrive at an estimate of $T_{\rm min} = 3020 \pm 850$\,K near 160\,\um\ for this star. This results in $T_{\rm min}/T_{\rm eff} = 0.58 \pm 0.17$, with $T_{\rm eff} = 5223 \pm 62$\,K. This value for \acenb\ is slightly lower than the optically derived value, but the error on the infrared ratio is large. However, also for the Sun itself, $T_{\rm min}$ determinations at different wavelengths show a wide spread of several hundred Kelvin \citep{avrett2003}.

In the present paper, we are however primarily concerned with the possible effects $T_{\rm min}$ might have on the estimation of very low emission levels from exo-Edgeworth-Kuiper belt dust. The intensity of the stellar model photosphere beyond 20 to 40\,\um\ is commonly estimated from the extrapolation of the spectral energy distribution (SED) into the Rayleigh-Jeans (RJ) regime at the  effective temperature $T_{\rm eff}$. There is a potential risk that this procedure will overestimate actual local stellar emissions, which may be suppressed at the lower radiation temperatures. In those cases, where the SEDs are seemingly well fit by the RJ-extrapolations, the differences may in fact be due to emission from cold circumstellar dust (exo-Edgeworth-Kuiper belts) and, here, we wish to quantify the magnitude of such an effect.

\subsection{Far Infrared: Detectability of cool dust from Edgeworth-Kuiper belt analogues} 

The SED of \acena\ is exceptionally well-determined over a broad range of wavelengths and we will use it here for the estimation of the likely level of this effect on other, non-resolved systems. We will do that by ``filling the pit with sand" to determine the corresponding fractional dust luminosity $f_{\rm d}\equiv L_{\rm dust}/L_{\rm star}$ and the accompanying temperature of the hypothetical dust.  

True Edgeworth-Kuiper belt analogues would be extremely faint and very difficult to detect \citep[$f_{{\rm d}\,\odot} \sim 10^{-7}$: ][]{teplitz1999,shannon2011,vitense2012}. For the estimation of faint exo-Edgeworth-Kuiper belt levels we apply simple dust ring models to ``fill" the flux-dip of the temperature minimum of \acena, i.e. by minimising the chi-square for 

\begin{equation}
\chi_\nu = \frac{S_{\rm dust}(\nu) - S_{\rm model}(\nu)}{\sigma}\,\,, \hspace{0.125cm} {\rm with\,limiting}\,\,\chi^2 \le 10^{-4}\,.
\label{fitgoodness}
\end{equation}

For illustrative purposes, we place the object at a distance of 10\,pc. Several ring temperatures were tested and dust fluxes were adjusted to match the stellar RJ-SED. The modified black body of the dust was then adjusted to coincide with the difference in flux (i.e. stellar black body minus measured flux) at 160\,\um\ and to a temperature that corresponds to the maximum of this flux difference.

This way of compensating for the $T_{\rm min}$-effect resulted in a dust temperature $T_{\rm dust} = 53$\,K at the modified black body radius of 34\,AU and a formal fractional luminosity $f_{\rm d}=(2.2^{+1.2}_{-1.5}) \times 10^{-7}$, almost one order of magnitude below the detection limit of {\it Herschel}-PACS at 100\,\um\ \citep[$f_{\rm d} \sim 10^{-6}$: see, e.g., Fig.\,1 of][]{eiroa2013} and comparable to that estimated for the solar system Edgeworth-Kuiper belt ($\sim 10^{-7}$, \citealt{vitense2012}). This fit is shown at the bottom of Fig.\,\ref{acenatminima} where it is compared to the stellar SED and the difference between the black body and the temperature minimum-fit, whereas the upper panel shows the observed data points, the adopted run of the $T_{\rm min}$-dip and that of the stellar RJ-SED. 

Based on simple arguments, this dust temperature provides also a first order estimate of the dust mass emitting at these wavelengths \citep[cf.][]{hildebrand1983}, viz. 

\begin{equation}
M_{\rm dust} \sim \frac { \left [ 1 -  \frac {S_{\rm min}(\nu)} {S_{\rm model}(\nu)}  \right ] S_{\rm model}(\nu)\,D^2  }  { \kappa_{\rm ext}(\nu)\,B(\nu,\,T_{\rm dust})  }
\label{mass}
\end{equation}

For an \acen-like star not showing the $T_{\rm min}$ effect, i.e. where the observed $S_{\rm obs,\,min}(\nu) = S_{\rm model}(\nu)$ (hence $M_{\rm dust}=0$), this formula would give a ``concealed" dust mass of the order of the solar Edgeworth-Kuiper Belt \citep[see, e.g., ][]{wyatt2008}, i.e.  \lapprox\,\powten{-3}\,$M_{\leftmoon}$ for $\kappa_{150\,\mu {\rm m}}$\,\gapprox\,$10\,{\rm cm}^2\,{\rm g}^{-1}$, depending on the largest size $a_{\rm max}$, the compactness and the degree of ice coating of the grains. For the cases considered here, reasonable values are within a factor of about 2 to 5 \citep[e.g., ][]{miyake1993,krugelsiebenmorgen1994,ossenkopf1994}. In general, $\kappa_{\rm ext}(\nu)=\kappa_{\rm sca}(\nu) + \kappa_{\rm abs}(\nu)$ is the frequency dependent mass extinction coefficient. At long wavelengths, say $\lambda > 5$\,\um, the scattering efficiency becomes negligible for the grains considered here and  $\kappa_{\rm ext}$ reduces to $\kappa_{\rm abs}$ ($\kappa_{\nu}$ henceforth).

In the far infrared, i.e. for low frequencies, one customlarily approximates the opacity by a power law, i.e.  $\kappa_{\nu}=\kappa_0 \left ( \nu/\nu_0 \right )^{\,\beta}$ when $\nu \le  \nu_0$, and where $\lambda_0 = c/\nu_0 > 2\pi a$ is a fiducial wavelength in the Rayleigh-Jeans regime, e.g. $\lambda_0=250$\,\um\  \citep{hildebrand1983}. For spherical Mie particles, the exponent of the frequency dependence of the emissivity $\beta$ is in the interval 1 to 2 for most grain materials \citep{emerson1988}. For blackbody radiation, $\beta=0$. For circumstellar dust, $\beta$-values around unity have commonly been found \citep[e.g.][]{krugelsiebenmorgen1994,beckwith2000,wyatt2008}.

The rough estimate from Eq.\,\ref{mass} refers only to the $a \le 1$\,mm-regime and will thus not account for larger bodies, in many cases dominating the mass budget of debris discs (as in the solar system). As is commonly done in debris disc studies, we distinguish between the dust mass and the total disc mass. To retrieve the mass of the unseen larger objects, numerical collisional models or analytical laws can be used \citep[e.g.,][]{thebault2007,wyatt2008,lohne2008,heng2010b,heng2010a,gaspar2012}, but we shall focus here solely on the observable dust mass, which can be reasonably derived from infrared and subillimetre observations \citep{heng2010b,heng2011}. Even when obtained with an expression as simple as Eq.\,\ref{mass}, these estimates are uncertain by at least a factor of several, because of uncertainties in the dust opacities.

From the dynamics of the binary system, we deduce (Appendix\,A) that a circumbinary disc is possible at a distance larger than $70 - 75$\,AU from the barycenter of the stars \citep[see also][]{wiegertholman1997,jaimeetal2012}. Comparing the synthetic image of such a ring with an assumed outer radius of 105\,AU in Fig.\,\ref{particleskyprojection} with our observed continuum maps (Figs.\,\ref{observations} and \ref{SPIRE_LABOCA}), we find seemingly coherent structures at the proper distances of a face-on circumbinary ring. It is important to note that the circumbinary's inclination may be unconstrained in contrast to the circumstellar discs which are dynamically limited to inclinations smaller than about 60\adeg, (\citealt{wiegertholman1997}; see also \citealt{moutou2011} for a more general spin-orbit-inclination study). Also, \citet{kennedy2012} reported a circumbinary and circumpolar dust disc around 99 Herculis which supports this possibility of non-coplanarity.

Regarding the possible detection of a circumbinary dust disc around \acena B, our single-epoch, multi-wavelength images would appear inconclusive. However, both proper motion and spectral line data obtained with APEX essentially rule out this scenario. If there really exists a circumbinary disc/ring around \acen tauri, it remained undetected by our observations.

\subsection{Mid Infrared: Warm zodi-dust in asteroid-like belts around \acena\ and \acenb}

\subsubsection{Stable orbits}

The tentative excesses at 24\,\um\ may be due to warm dust.  However, the binary nature of the \acen\ system limits the existence of stable orbits to three possibilities. These are one large circumbinary disc with a certain inner radius and two circumstellar discs with maximum (or hereafter, critical) semi-major axes. These critical semi-major axes can be found using the semi-analytical expression of \citet{holmanwiegert1999}, viz.

\begin{equation}
a_{\rm crit} = \left( c_1 + c_2 \, \mu + c_3 \, e + c_4 \, \mu \, e + c_5 \, e^2 + c_6 \, \mu \, e^2 \right) \, a_{\rm AB}
\label{critaxis}
\end{equation}
where $e$ is the eccentricity of the binary orbit, $\mu = M_{\rm B} /(M_{\rm A} + M_{\rm B}$ i.e. the fractional mass of the stars, $a_{\rm AB}$ is the semi-major axis of the binary's orbit, and the coefficients $c_1$ through $c_6$ (all with significant error bars) were computed by \citet{holmanwiegert1999}.

Using known parameters for \acen\ (see Table\,\ref{starprop}) we can deduce that the circumstellar discs can not be larger than $2.78 \pm 1.48\,$AU around \acena\ and $2.52 \pm 1.60\,$AU around \acenb, i.e. smaller than $\sim 4\,$AU around either star (Table\,\ref{starprop}). 

The putative Earth-mass planet, \acenb b \citep{dumusque2012} is small enough ($\sim 1.13$\,\mearth) that its Hill radius is just $4 \times 10^{-4}$\,AU. Our model discs never reach closer to the star than 0.08\,AU and, consequently, this planet was neglected in our disc modelling. 

\begin{table}
    \caption{Properties of the \acen tauri binary}
    \label{starprop}
    \begin{tabular}{lcc}
\noalign{\smallskip}
    \hline \hline 
\noalign{\smallskip}
     & $\alpha$ Cen A & $\alpha$ Cen B \\
\noalign{\smallskip}
    \hline
\noalign{\smallskip}
    Sp.Type$^a$ 			   	& G2 V                      		& K1 V \\
    $T_{\rm eff}$ (K)$^b$      		& $5824 \pm 24$             	& $5223 \pm 62$ \\
    $L_{\rm star}$ (\lsun)$^b$ 	& $1.549^{+0.029}_{-0.028}$ 	& $0.498^{+0.025}_{-0.024}$ \\
    $M_{\rm star}$ (\msun)$^{bc}$  	& $1.105 \pm 0.007$         & $0.934 \pm 0.006$ \\
    $R_{\rm star}$ (\rsun)$^a$     	& $1.224 \pm 0.003$         & $0.863 \pm 0.005$ \\
    $a_{\rm crit}$ (AU)$^{d}$		& $2.778 \pm 1.476$         & $2.522 \pm 1.598$ \\
\noalign{\smallskip}
    \hline
\noalign{\smallskip}
    \multicolumn{3}{c}{Common parameters} \\
\noalign{\smallskip}
    \hline
\noalign{\smallskip}
    \multicolumn{2}{l}{Inclination to LOS, $i$ ($\degr$)$^c$}             	& $79.20 \pm 0.04$ \\
    \multicolumn{2}{l}{Arg. of periapsis, $\omega$ ($\degr$)$^c$}  	& $231.65 \pm 0.08$ \\
    \multicolumn{2}{l}{Long. of asc. node, $\Omega$ ($\degr$)$^c$} 	& $204.85 \pm 0.08$ \\
    \multicolumn{2}{l}{Period (yr)$^{bc}$}                        			& $79.91 \pm 0.01$ \\
    \multicolumn{2}{l}{Eccentricity$^c$}                           			& $0.5179 \pm 0.0008$ \\
    \multicolumn{2}{l}{Distance (pc)$^b$}                          			& $1.348 \pm 0.035$ \\
    \multicolumn{2}{l}{Age (yr)$^e$}                               				& $(4.85 \pm 0.50) \times 10^9$ \\
\noalign{\smallskip}
    \hline
    \end{tabular}
    \begin{list}{}{}
        \item[$^a$] \citet{kervella2003}, $^b$ \citet{torresetal2010}, $^c$ \citet{pourbaixetal2002}, $^d$ \citet{holmanwiegert1999}, $^e$ \citet{thevenin2002}.
    \end{list}    
\end{table}

\begin{figure}
  \begin{center}
    \includegraphics[width=88mm]{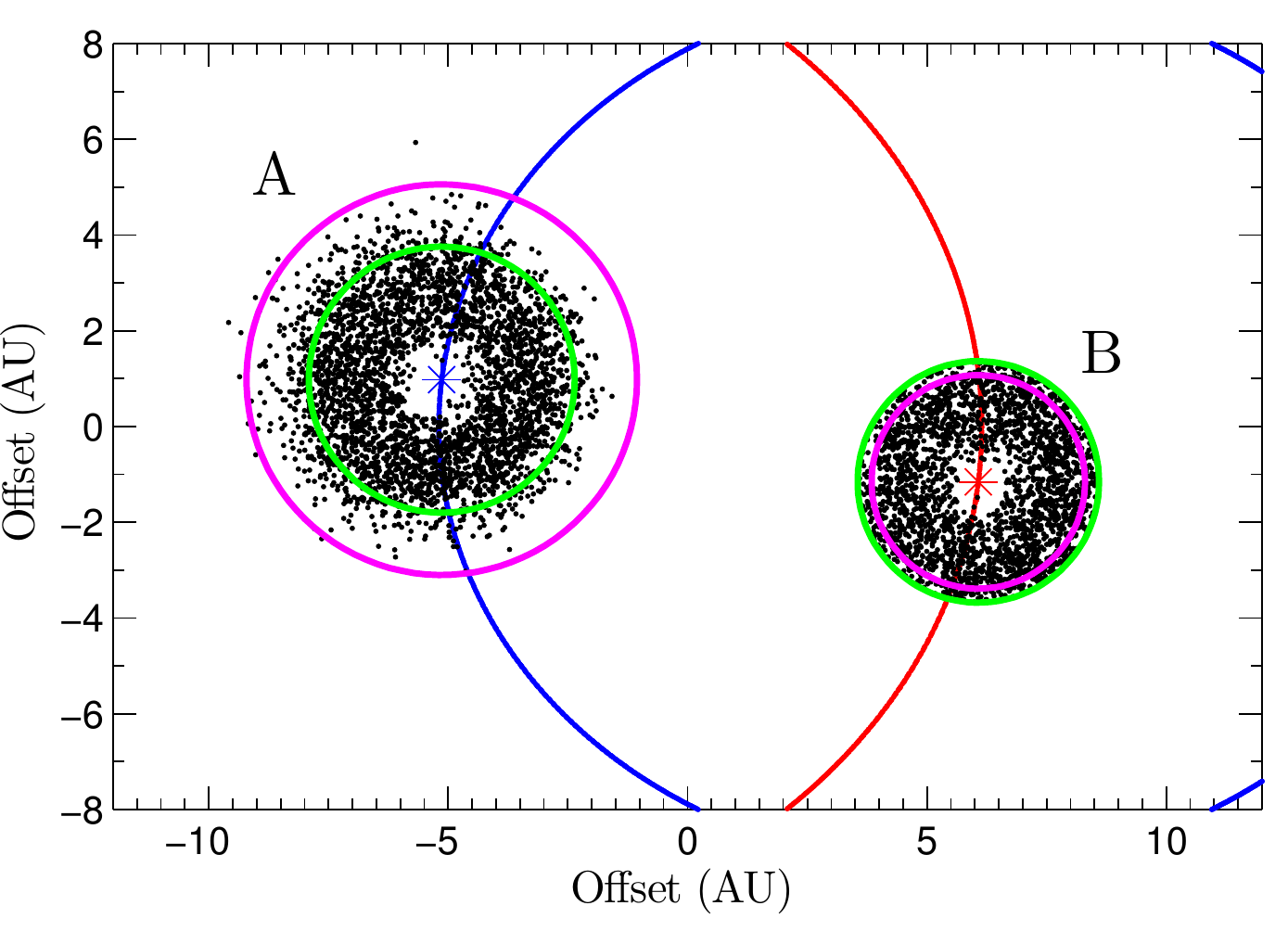}
    \caption{Face-on circumstellar test-particle discs after $\sim 10^3$ periods shown with the stars close to periapsis. \acena\ is colour coded blue (the left star and its orbit) and \acenb\ is colour coded red (the right star and its orbit). The green circles represent $a_{\rm crit}$ around the stars and the magenta circles show estimates of  their respective snow lines.}
	\label{circstfaceon}
  \end{center}
\end{figure}

Due to the large errors of the estimated $a_{\rm crit}$ it was useful to test these limits with simple test-particle simulations of mass- and size-less particles. The resulting particle discs will also be used as a basis for radiative transfer simulations later. A steady-state of these are shown in Fig.\,\ref{circstfaceon} where the accuracy of the \citet{holmanwiegert1999} estimates is clear. Each simulation run was left for $10^3$ orbital periods (i.e. $\sim 8 \times 10^4$\,yr). The dynamics had earlier been examined also by others \citep[e.g., ][]{benest1988, wiegertholman1997,holmanwiegert1999,lissauer2004,thebault2009}.

These size limits are reminiscent of the inner solar system, i.e. this opens the possibility for an asteroid belt-analogue for each star which forms dust discs through the grinding of asteroids and comets. Temperature estimates for the solar system zodiacal cloud are around 270\,K and that of the fractional luminosity is about $10^{-7}$ \citep{fixsen2002,nesvorny2010,roberge2012}, i.e. $f_{\rm d}$ is of similar magnitude as that of the Edgeworth-Kuiper belt \citep{vitense2012}.

\subsubsection{Disc temperatures and snow lines}

The {\it Spitzer} observations at 24\,\um\ \citep[PSF\,$\ge 6$\asec, Table\,\ref{fluxes};  see also  ][]{kennedy2012} would not fully resolve these dynamically allowed discs. Strict lower temperature limits are provided by black body radiation from the discs, i.e. $T_{\rm bb} = \sqrt{R_{\rm star} /(2\,R_{\rm bb})}\,T_{\rm eff} =186$\,K and 147\,K respectively for \acena\ and B, where the ratio of absorption to emission efficiency of unity for a black body has been used \citep[e.g., ][]{liseauetal2008,lestrade2012,heng2013}.

The \acen\ circumstellar discs are constrained to radii smaller than 4\,AU and should be relatively warm. We need to know, therefore, the locations of the snow lines in order to understand the likelihood of the presence of icy grains on stable orbits around these stars.

The snow line can be defined to be the largest radial distance at which the sublimation time scale is larger than all other relevant time scales of the system \citep{artymowicz1997}. Our estimations are based on the equations and methods by \citet{grigorieva2007} and by \citet{lamy1974}. The sublimation time was compared with the orbital time as a lower limit and the system age ($4.85 \times 10^9\,$yr) as an upper limit. For the temperature-radius relation we assumed blackbody emitters. This analysis resulted in sublimation temperatures between 154\,K and 107\,K  for \acena, at the radial distances of 4.08 and 8.48\,AU, respectively. Therefore, the snow line of \acena\ seems to be outside its dynamically stable region of 2.78\,AU, which makes the presence of any icy disc grains not very likely (Fig.\,\ref{circstfaceon}).

For \acenb, sublimation temperatures were found to be between 157\,K and 107\,K, which corresponds to 2.23 and 4.81\,AU, respectively. The lower bound is well inside the critical semi-major axis of 2.52\,AU (see Fig.\ref{circstfaceon}). As this is based on an initial grain size of 1\,mm (smaller sizes move the snow line outward, while larger ones move it inward), we can assume that the existence of a ring of larger icy grains and planetesimals is possible at the outer edge of the \acenb\ disc. Such a ring could supply the disc with icy grains. However, these can not be expected to survive in such a warm environment for long. Thus we will assume that the iceless opacities from \citet{miyake1993} are sufficient in this case. \citet{ossenkopf1994} computed opacity models for the growth of ice coatings on grains. This conglomeration results generally in larger opacities than for their bare initial state. 

We use {\it dust} opacities that are based on the models presented by \citet{miyake1993}, but scaled by a gas-to-dust mass ratio of one hundred. For a ``standard" size distribution \citep{lestrade2012}, i.e. $n(a) \propto \,a^{-3.5}$, the work by \citet{ossenkopf1994} gave similar results \citep[see also, e.g., ][]{krugelsiebenmorgen1994,stognienko1995,beckwith2000}. These works had different scopes, addressing specifically coagulation processes, rather than destructive collisions. However, for bare grains these opacites may to some extent be applicable also for the dust in debris discs (see also the home page of B.\,Draine, {\tiny \texttt{http://www.astro.princeton.edu/$\sim$draine/dust/dustmix.html}}). 

In the following sections we present synthetic SEDs obtained with the radiative transfer program RADMC-3D \citep{dullemond2012} to assess the disc configurations in more detail.

\subsection{Energy balance, temperature and density distributions}

Isothermal blackbody emission from thin rings would not make a realistic scenario for the emission from dust belts or discs. In particular, we need to specify the run of temperature and density, which is obtained from the solution to the energy equation, balancing the radiative heating with the cooling by the optically thin radiation.  

In this balance equation, the loss term expresses the flux density of this thermal dust emission from an ensemble of grains received at the Earth, viz.

\begin{equation}
S_{\rm dust}(\nu) = \int \int 4 \pi a^2 Q_{\nu,\,{\rm abs}}(a) \frac{ \pi B_{\nu}(T, {\bf r})}{4 \pi D^2} n(a,{\bf r})\,da\,d{\bf r}^3
\label{cooling}
\end{equation}

where $4 \pi a^2 Q_{\nu,\,{\rm abs}}(a)$ is the emissivity of the spherical grains with radius $a$, and where $Q_{\nu,\,{\rm abs}}(a)$ is the absorption coefficient. $n(a,{\bf r})$ is the volume density of the grains at location ${\bf r}$. $\pi B_{\nu}$ is the thermal emittance at the equilibrium temperature $T$ at {\bf r}, and $D$ is the distance to the source (so that $4 \pi D^2 \int S_{\rm dust}(\nu)\,d\nu = L_{\rm dust}$). The absorption coefficient $Q_{\nu,\,{\rm abs}}$ is related to the dust opacity through $Q_{\nu,\,{\rm abs} } =  (4/3)\,\kappa_{\nu}\,\rho\,a$, and where $\rho$ is the grain mass density. $Q_{\nu,\,{\rm abs}}$ is shown as function of wavelength for a variety of materials and compositions for two grain sizes (10 and 1000\,\um)  in Fig.\,4 of \citet{krivov2013}. We shall discuss grain sizes in the next section.

\subsubsection{Grain size distribution}

It is often assumed that $n(a)\, \propto\, a^{\,q}$ describes the distribution in size of the grains \citep[e.g., ][]{miyake1993,krugelsiebenmorgen1994,ossenkopf1994,krivov2000}. However, both the shape and the existence and magnitude of the power law exponent $q$ have been a matter of intense debate since Dohnanyi's classic work on interplanetary particles in the solar system \citep{dohnanyi1969}. For a collision dominated system at steady state, his result is summarised in a nutshell by the parameter $s=11/6$, implying that $q = 2-3 s = - 3.5$ \citep[see, e.g., ][]{dominik2003,wyatt2008}. This original work referred to the zodiacal cloud, but more recent estimates also for the much larger ``grains", i.e. Edgeworth-Kuiper belt objects (TNOs), by \citet{bernstein2004} lead to a similar frequency spectrum for the diameters, $dn/dD\,\propto\,D^{-p}$ and $p=4\pm 0.5$, so that $q = -3 \pm 0.5$ \citep[see also][]{fraser2009}.

More recent works have been following the collisional time evolution of dusty debris discs using numerical statistical codes. They have found rather strong deviations from true power laws, showing wiggles and wavy forms due to resonances \citep{krivov2006,thebault2007,lohne2008,kral2013,krivov2013}. However, mean values of these oscillations may still be consistent with power law exponents which are not too dissimilar from the steady state Dohnanyi-distribution. In fact, \citet{gaspar2012} found from a large number of simulations that $q \sim -3.65$, i.e. slightly steeper than commonly assumed. Quoting \citet{miyake1993}, ``$q$ will be large ($\sim -2.5$) if the coagulation processes are dominant, whereas $q$ will be small ($\sim -3.5$) if the disruption processes are dominant", one might infer that disruptive collisions would dominate in debris discs, as expected. To be consistent with the exploited opacities, but also for comparison reasons, we are using a Dohnanyi distribution, i.e. $n(a)\,\propto\,a^{-3.5}$ which we will use henceforth. For simplicity, the distribution of particle sizes in the disc is assumed homogeneous\footnote{In general, grain sizes may not be homogeneously distributed throughout circumstellar discs. Smaller grains may be more abundant in the outer parts of the discs and may even reside in the dynamical unstable regions in a binary, due to production through collisions and radiation pressure (e.g. \citealt{thebaultetal2010}).} as the grains are large enough to not be significantly affected by drag forces and radiation pressure.
 
In Eq.\,\ref{cooling}, the integral over $a$ extends over the finite range $a_{\rm min}$ to $a_{\rm max}$, where we adopt for $a_{\rm max}$ the upper bound for the $\kappa_{\nu}$-calculations, i.e. 1\,mm. The smallest size, $a_{\rm min}$, is found for grains large enough not to be susceptible to the combined radiation pressure and stellar wind drag \citep{strubbe2006,plavchan2009}, i.e.

\begin{equation}
a_{\rm min} > a_{\rm blow} \approx \frac{ 3 }  { 8\,\pi\,G\,M_{\rm star}\,\rho_{\rm grain} } \left [ \frac {L_{\rm star}} {c} +  \dot{M}_{\rm star}\, \upsilon_{\rm w} \right ]\,\,,
\label{blowout}
\end{equation}

where we, for this order of magnitude estimate, have set the radiative and wind coupling coefficients to unity.  Inside the brackets are the radiative and the mechanical momentum rates (``forces"), respectively. The stellar mass loss rate of \acena\ is of the order of that of the Sun, i.e. $2\times 10^{-14}$\,\msunyr, and also the average wind velocity is similar, viz. $\upsilon_{\rm w} \sim 400$\,\kms\ \citep{wood2001,wood2005}\footnote{Note that in this work the binary is not resolved into its components A and B, so that the quoted values refer to both of them. Athough \acen\,B is the more active star of the two, we simply assumed equal mass loss rates.}. This implies that the second term in Eq.\,\ref{blowout} is negligible and that $a_{\rm blow} = 0.64$\,\um\ for the stellar parameters of Table\,\ref{starprop} and a grain density of 2.5\,g\,\cmthree\ (for \acenb, the corresponding $a$-value is about three times lower). It has been shown that the lower cut-off size is smooth \citep[e.g.][and references therein]{wyatt2011,loehne2012} and up to about six times the blow-out radius, i.e. $a_{\rm min} \sim 4$\,\um. 

In the next section, we will calculate a number of numerical models for Eq.\,(\ref{cooling}) by varying parameter values to get a feeling for how well (and uniquely) the observations can be reproduced. There were marginal excesses corresponding to $2.5\,\sigma$ at 24\,\um\ for both \acena\ and \acenb.
 
\begin{figure*}
  \begin{center}
    \includegraphics[width=180mm]{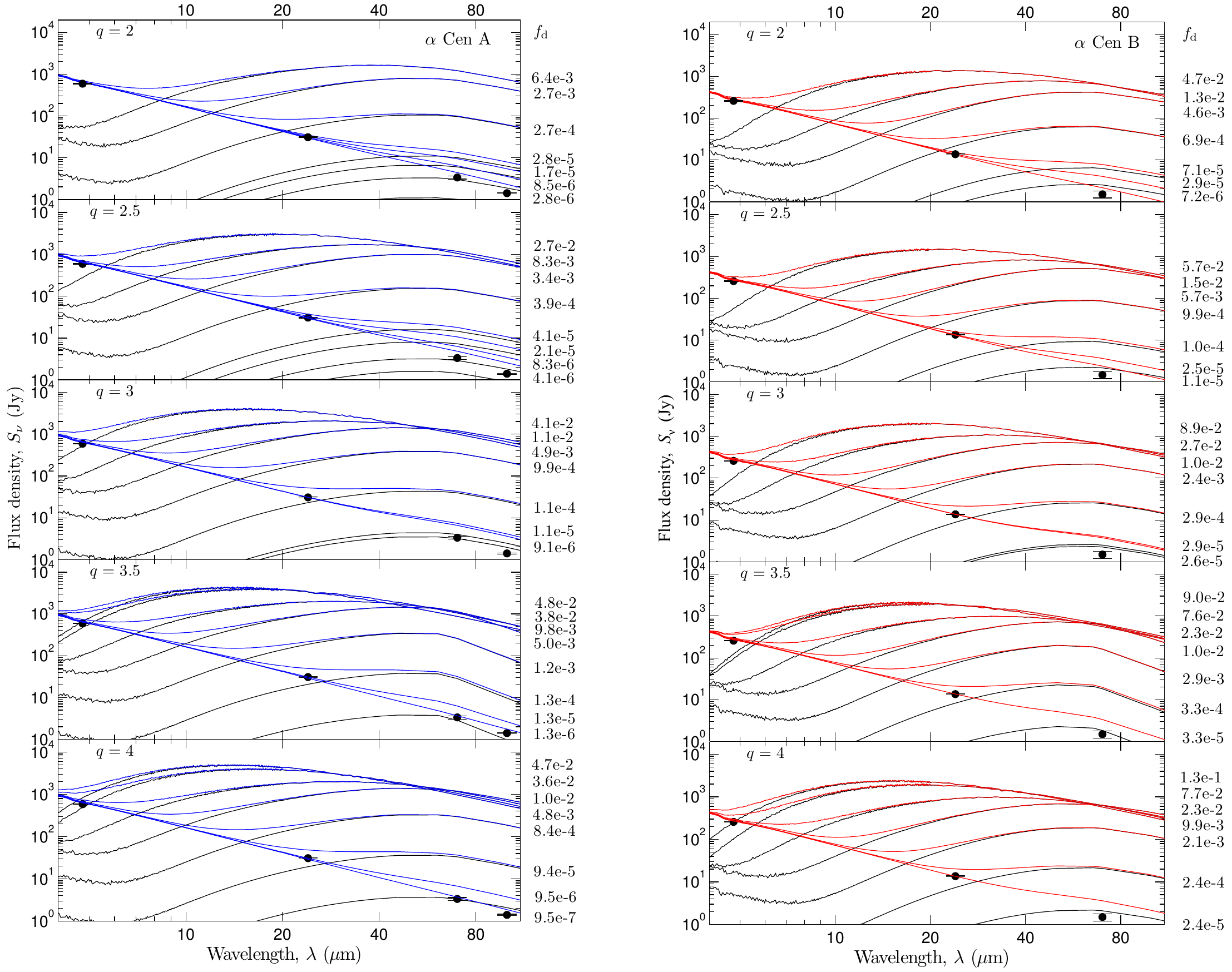}
    \caption{{\bf Left:} Mid-infrared SEDs for \acena\ for different values of the power law exponent $q$ of the size distribution function, viz. $n(a)\,\propto\,a^{\,-q}$. Black curves depict dust SEDs, whereas blue curves display the total, i.e. $S_{\nu}^{\rm star}+S_{\nu}^{\rm dust}$. For each $q$, there are graphs for a selection of initial dust mass estimates and the fractional dust luminosity, $f_{\rm d}= L_{\rm dust}/L_{\rm star}$, is shown next to each curve. The 100\,\um\ data were not part of the fitting procedure. {\bf Right:} Same as the left panel but for \acenb\ and with the total emission shown in red. }
	\label{q-distribution}
  \end{center}
\end{figure*}
 
\subsubsection{RADMC-3D exozodi-disc calculations}

RADMC-3D is a Monte-Carlo radiative transfer code, written primarily for continuum radiation through dusty media in an arbitrary three-dimensional geometry (developed by \citealt{dullemond2012})\footnote{\tiny{\texttt{http://www.ita.uni-heidelberg.de/$\sim$dullemond/software/ \\RADMC-3D/}}}. It might seem an enormous ``overkill" to use a radiative transfer program for the optically very thin discs discussed here, where the optical depth $\tau$ is comparable to the fractional luminosity $f_{\rm d}$ of the dust disc. However, RADMC-3D is a convenient and user-friendly means to solve the energy equation, calculate SEDs, and convolve the model-images with the appropriate response functions of the used equipment for the observations (e.g., transmission curves, PSFs or beams, etc.). Details regarding the computational procedure can be found in Appendix\,A and examined parameters are presented in Table\,\ref{params}.

The basic input to RADMC-3D is a stellar model atmosphere (we use three-dimensional interpolations in the PHOENIX/GAIA grid, \citealt{brott2005}), a grid with dust densities for each included dust species, and absorption and scattering coefficients at each wavelength and for each dust species. The inner disc radii are given by the dust vaporation temperature\footnote{This corresponds approximately to the low-density extrapolation of the data for olivine given by \citet{pollack1994}.} of  \about\,750\,K and the outer radii determined by the binary dynamics. The discs have assumed height profiles $h \sim 0.05\,r^{\,\,\beta}$, with $\beta=0$  \citep[or $\beta=1$; e.g.,][]{wyatt2008}.  

Also based on the dynamical model parameters are the dust mass estimations for various power law exponents of the surface density, $\Sigma\,\propto\,r^{-\gamma}$, where $\gamma \ge 0$. For reference, the Solar Minimum Mass Nebula (SMMN) has $\gamma = 1.5$ \citep{hayashi1981}. For the SMMN, at $r_0 = 1 $\,AU, $\Sigma_0 \sim 2\times 10^3$\,g\,\cmtwo, whereas exo-planet systems may have slightly steeper laws and somewhat higher $\Sigma_0$ \citep{kuchner2004}.  

Various grain size distributions were examined in accordance with Fig.\,7 of \cite{miyake1993}, i.e.  $-4 \le q \le -2$. The results of these computations are visualized in Fig.\,\ref{q-distribution}. There, we investigate if any combination of parameters could be consistent with the observed data, i.e. the flux densities at 5, 24 and 70\,\um. For \acena, acceptable models  would imply values of  $f_{\rm d} \sim 10^{-5}$ for its hypothetical zodi, where $q = -4$ to $-3.5$. This would appear to be in agreement with the result by \citet{gaspar2012}, i.e. $q=-3.65$. 

\begin{figure}
  \begin{center}
    \includegraphics[width=88mm]{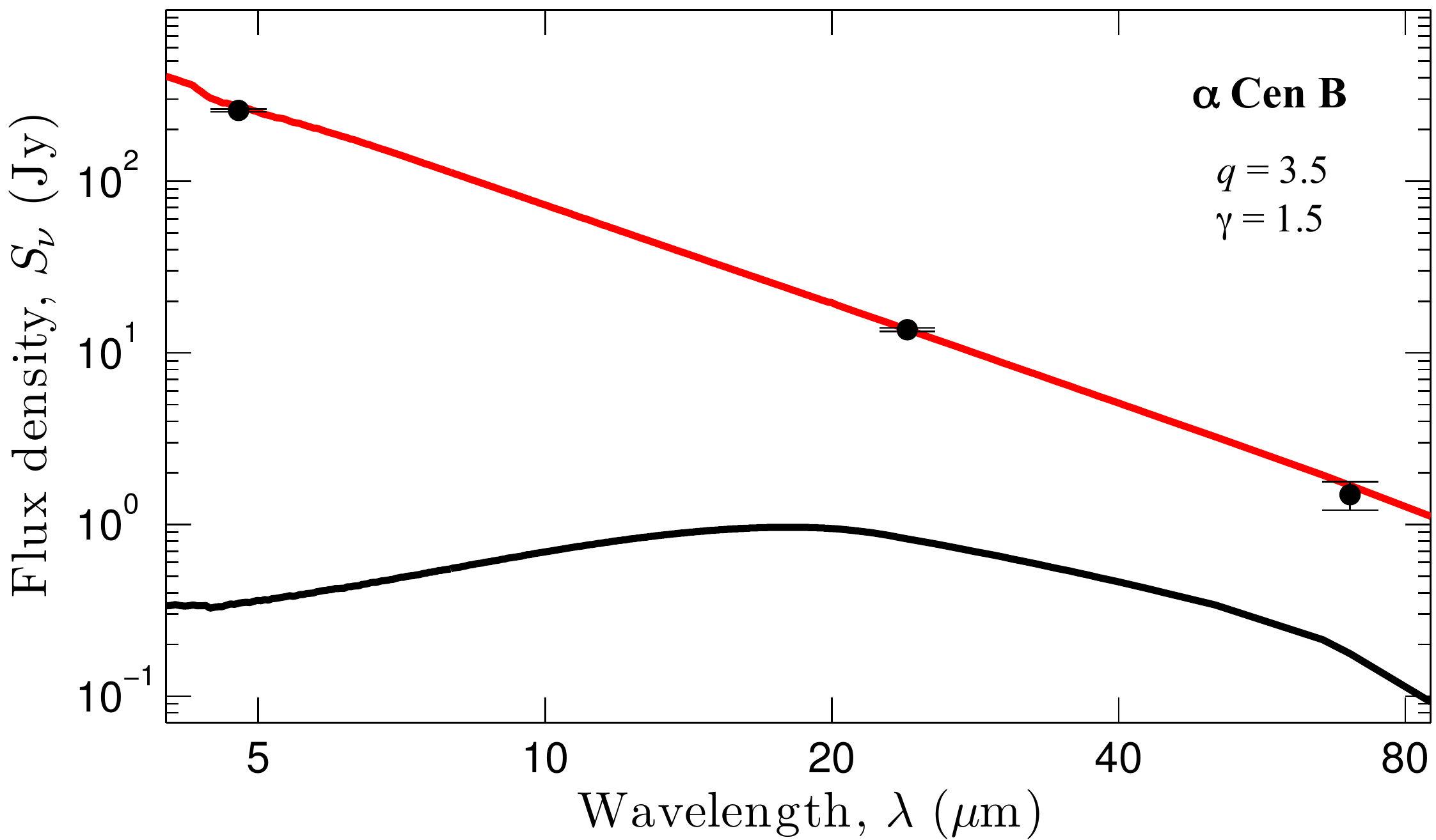}
    \caption{Mid-infrared SED of  \acenb\ (star + dust, red curve) with data points. Assuming a ``standard" disc model with $q=3.5$ and $\gamma=1.5$, the black curve shows the maximum emission from circumstellar dust that is consistent with the data.}
	\label{standard}
  \end{center}
\end{figure}

However, no such model was found that would fit the data for \acenb, as illustrated in the right hand panel of Fig.\,\ref{q-distribution}. These models assume a flat $\Sigma$-distribution, i.e. $\gamma=0$. To be compatible with the observations, models for $\gamma \neq 0$ would require very large values of $\gamma$ (up to 10), making these ``discs" more ring-like. However still with $\gamma=0$, best agreement was achieved, when the model disc of \acenb\ had smaller extent than the dynamically allowed size, i.e. $R_{\rm in}=0.1$\,AU and $R_{\rm out}=0.5$\,AU. The temperature profile is given by $T(r) = 745\,(r/0.05\,{\rm AU})^{-0.55}$\,K and the dust-SED results in an $f_{\rm d}=3\times 10^{-5}$ for a dust mass $m_{\rm d}=4\times 10^{-6}\,M_{\leftmoon}$ ($3\times 10^{20}$\,g, i.e. about $10-30\,m_{Z_{\odot}}$), for grains with $a= 4$\,\um\ to 1\,mm and with sizes according to a $q=-3.5$ power law (see Fig.\,\ref{standard}).  

For both \acena\ and B, these limiting values of $f_d$ and $m_d$, respectively, would correspond to some $10^2$ and 10 times those of the solar system zodi dust in \lapprox\,100\,\um\ size particles \citep{roberge2012,fixsen2002,nesvorny2010}. The latter authors estimate the total mass of the asteroid belt at about $0.03\,M_{\leftmoon}$ ($2\times 10^{24}$\,g). \citet{heng2010b, heng2010a} have calculated total disc masses for ages of about 3\,Gyr. Their Fig.\,7 would indicate that the mass of a ``collision-limited" disc with a semimajor axis \lapprox\,4\,AU and detected at 24\,\um\ would amount to $\sim 0.1\,M_{\leftmoon}$. In both cases, these estimations refer effectively to discs around single stars and it is not entirely clear to what degree this applies also to multiple systems.

\begin{table}
    \caption{Model parameters for RADMC-3D disc calculations.}
    \label{params}
    \begin{tabular}{lc}
\noalign{\smallskip}
    \hline \hline 
\noalign{\smallskip}
  Parameter (unit)								& Value 	\\  
\noalign{\smallskip}
    \hline
\noalign{\smallskip}
lower grain cut-off size, $a_{\rm min}$ (\um)   			& 4 		\\
upper grain size limit,    $a_{\rm max}$ (\um)   			& 1000 	\\
power law exponent,     $n(a)\,\propto\,a^{\,-q}$	 (\cmthree)& 2.0, 2.5, 3.0, 3.5, 4.0		\\
power law exponent,     $h(r)\,\propto\,r^{\,\beta}$ (AU)	& 0.0, 1.0		\\
power law exponent, $\Sigma(r)\,\propto\,r^{\,-\gamma}$ (g\,\cmtwo)& 0, 1/2, 1, 3/2, 2, 6, 10 \\
grain mass density, $\rho$ (g\,\cmthree) 				& 2.5		\\ 
vaporization temperature, $T_{\rm vap}$ (K)			& 750	\\
inner disc radius, $R_{\rm in}$ 	(AU)					& $r(T_{\rm vap})$	\\
outer disc radius, $R_{\rm out}$ (AU)				& $a_{\rm crit}$		\\
logarithmic dust mass ($M_{\leftmoon}$)		 		& $-7$ through $+4$ by 0.1 \\
\noalign{\smallskip}
    \hline \hline
 \end{tabular}
\end{table}

\section{Conclusions}

Based on FIR and submm photometric imaging observations from space ({\it Herschel})  and ground (APEX) of \acen\ the following main conclusions can briefly be summarised as

\begin{itemize}
\item[$\bullet$]	both binary components A and B were detected individually at 70\,\um. At longer wavelengths, $100-870$\,\um, the pair was unresolved spatially during the time of observation, i.e. \asecdot{8}{4} in 2007 to  \asecdot{5}{7} in 2011; 
\item[$\bullet$] at 100 and 160\,\um, the solar twin \acena\ displays a temperature minimum in its SED, where the radiation originates at the bottom of its chromosphere. At longer wavelengths higher radiation temperatures are observed. This $T_{\rm min}$-effect is also marginally observed for \acenb;
\item[$\bullet$] for solar-type stars not displaying the $T_{\rm min}$-phenomenon, compensating for the ``missing flux" by radiation from optically thin dust could account for fractional dust luminosities, $f_{\rm d} \sim 2 \times 10^{-7}$, comparable to that of the solar system Edgeworth-Kuiper belt;
\item[$\bullet$] combining {\it Herschel} data with mid-IR observations ({\it Spitzer}) indicates marginal ($2.5\,\sigma$) excess emission at 24\,\um\ in the SEDs of both stars. If due to circumstellar emission from dust discs, fractional luminosity and dust mass levels would be some 10 to 100 times those of the solar zodiacal cloud.
\end{itemize}

\begin{acknowledgement}
We thank the referee for his/her critical reading of the manuscript and the valuable suggestions which improved the quality of the paper. We are also grateful to H.\,Olofsson to grant his {\it Director's Discretionary Time} to this project. We also wish to thank P.\,Bergman for his help with the APEX observations on such short notice and the swift reduction of the data. We appreciate the continued support of the Swedish National Space Board (SNSB) for our {\it Herschel}-projects. The Swedish authors appreciate the continued support by the Swedish National Space Board (SNSB) for our {\it Herschel}-projects. C.\,Eiroa, J.P.\,Marshall, and B.\,Montesinos are partially supported by Spanish grant AYA 2011/26202. A.\,Bayo was co-funded under the Marie Curie Actions of the European Comission (FP7-COFUND). S.\,Ertel thanks the French National Research Agency (ANR) for financial support through contract ANR-2010 BLAN-0505-01 (EXOZODI). 

\end{acknowledgement}

\def\apjl{ApJL}
\def\aj{AJ}
\def\apj{ApJ}
\def\pasp{PASP}
\def\spie{SPIE}
\def\apjs{ApJS}
\def\araa{ARAA}
\def\aap{A\&A}
\def\nat{Nature}
\def\mnras{MNRAS}

\bibliographystyle{aa} 
\bibliography{references_phd} 

\appendix
\section{Orbital dynamics and disc modelling}

The test particle simulations were done for two reasons; firstly, to assess the binary dynamics per se, and, secondly, for the steady state results to be used as disc models for radiative transfer simulations.

The standard equation of motion for circumstellar dust grains (around a single star) sums up most of the physical processes a grain experiences and is written \citep{robertson1937,starkkuchner2008}

\begin{equation}
\begin{split}
\vec a_n = & - \frac{G \, M_\star}{r_n^3} \, (1 - \mathcal{B}_n) \, \vec r_n \\
           & - \frac{(1 + SW) \, \mathcal{B}_n}{c} \, \frac{G \, M_\star}{r_n^2} \, (\dot r_n \, \hat r + \vec v_n) \\
           & + \displaystyle\sum_i \, \frac{G \, m_i}{|\vec r_i - \vec r_n|^3} \, (\vec r_i - \vec r_n)
\end{split}
\label{standardgrainmotion}
\end{equation}
for each test particle $n$. Here $\mathcal{B}_n$ is the ratio between radiation pressure and gravitation exhibited on particle $n$, $SW$ is the ratio between stellar-wind drag and Poynting-Robertson drag and is for the Sun about one third \citep [\about\,0.2--0.3, ][]{gustafson1994}, and $m_i$ and $\vec r_i$ is the mass and position of any planet in the system.

As hinted at in Sect.\,4.5.1, we consider grain sizes $\ge 4$\,\um. This is more than six times the blow-out radius and, consequently, we only solve the classical restricted three-body-problem, i.e.

\begin{equation}
\vec a_n = - \frac{G \, M_{\rm A}}{|\vec r_n - \vec r_{\rm A}|^3 } \, (\vec r_n - \vec r_{\rm A}) - \frac{G \, M_{\rm B}}{|\vec r_n - \vec r_{\rm B}|^3 } \, (\vec r_n - \vec r_{\rm B})
\label{ourgrainmotion}
\end{equation}
for each test particle $n$. These simulations implemented a Runge-Kutta 4 integrator to solve the motions.

One particle disc was simulated at a time. The two circumstellar discs had identical initial conditions except for the central star, i.e. they had $10^4$ particles in a disc with constant particle density and smaller than 5\,AU radius. The circumbinary disc also had constant particle density but was 100\,AU in radius (this was varied, but we only show results from the disc with 100\,AU radius). The circumbinary disc's outer radius is, in contrast to the circumstellar discs, not constrained by the employed dynamics. Therefore the outer edge seen in Fig.\,\ref{particleskyprojection} is artificial.

All discs were left to run for $10^3$ periods, i.e. $\sim 8 \times 10^4\,$yr which was found to be sufficient to achieve a steady state. That the disc had settled by that time was verified with much longer runs (e.g., $\sim 8 \times 10^5\,$yr). At the inner edge of the circumstellar discs, the grid resolution was up to 0.047\,AU. On the much larger scaleof the circumbinary disc, time steps of  0.1\,yr were sufficient. 

Finally, particles that came either closer than 0.001 Hill radii of each star or farther out than 1000\,AU from the barycenter were removed. These limits were arbitrarily set on the basis of previous tests. The number of particles remaining varied from run to run with e.g. 2496 as the lowest (\acenb\ with initial disc radius of 5\,AU) up to 6717 (again \acenb\ but with initial disc radius of 0.2\,AU). The \acena\ disc with initial radius of 5\,AU had 3538 remaining particles.

To compute dust SEDs the dust disc density grid was obtained directly from the steady state discs. These were converted to a grid of mass densities with a large number of initially guessed dust masses, $m_{\rm dust}$. The proper temperature and density profiles yielded SEDs which were compared to the data. The best fit was selected on the basis of its $\chi^2$-value.

\end{document}